\documentclass[english]{article}
\usepackage[T1]{fontenc}
\usepackage[latin9]{inputenc}
\usepackage{babel}
\usepackage{array}
\usepackage{float}
\usepackage{amsmath}
\usepackage{amssymb}
\usepackage{graphicx}
\usepackage[unicode=true,
bookmarks=false,
breaklinks=false,pdfborder={0 0 1},colorlinks=false]
{hyperref}
\makeatletter

\providecommand{\tabularnewline}{\\}
\floatstyle{ruled}
\newfloat{algorithm}{tbp}{loa}
\providecommand{\algorithmname}{Algorithm}
\floatname{algorithm}{\protect\algorithmname}

\usepackage{babel}
\usepackage{algpseudocode}
\usepackage{algorithm}
\usepackage{algorithmicx}
\usepackage{mathrsfs}

\algrenewcommand\algorithmicrequire{\textbf{Input:}}
\algrenewcommand\algorithmicensure{\textbf{Output:}}
\algdef{SE}[DOWHILE]{Do}{doWhile}{\algorithmicdo}[1]{\algorithmicwhile\ #1}%
\newcommand{\bs}{\boldsymbol}

\usepackage{lineno}

\@ifundefined{showcaptionsetup}{}{%
 \PassOptionsToPackage{caption=false}{subfig}}
\usepackage{subfig}
\makeatother

\begin{document}

\title{Hyper-reduction over nonlinear manifolds for large
nonlinear mechanical systems}

\author{Shobhit Jain, Paolo Tiso}
\date{\vspace{-5ex}}
\author{Shobhit Jain\footnote{Corresponding author, E-mail: shjain@ethz.ch, Phone: +41 44 632 77 55}, Paolo Tiso}
\maketitle
\begin{center}
	Institute for Mechanical Systems, ETH Zürich \\
	Leonhardstrasse 21, 8092 Zürich, Switzerland
	\par\end{center}

\begin{abstract}
	Common trends in model reduction of large nonlinear finite-element-discretized systems involve Galerkin projection of the governing equations onto a low-dimensional linear subspace. Though this reduces the number of unknowns in the system, the computational cost for obtaining the reduced solution could still be high due to the prohibitive computational costs involved in the evaluation of nonlinear terms. Hyper-reduction methods are then used for fast approximation of these nonlinear terms. In the finite-element context, the energy conserving sampling and weighing (ECSW) method has emerged as an effective tool for hyper-reduction of Galerkin-projection-based reduced-order models (ROM).
	
	More recent trends in model reduction involve the use of nonlinear manifolds, which involves projection onto the tangent space of the manifold. While there are many methods to identify such nonlinear manifolds, hyper-reduction techniques to accelerate computation in such ROMs are rare. In this work, we propose an extension to ECSW to allow for hyper-reduction using nonlinear mappings, while retaining its desirable stability and structure-preserving properties. As a proof of concept, the proposed hyper-reduction technique is  demonstrated over models of a flat plate and a realistic wing structure, whose dynamics have been shown to evolve over a nonlinear (quadratic) manifold. An online speed-up of over one thousand times relative to the full system has been obtained for the wing structure using the proposed method, which is higher than its linear counterpart using the ECSW.
\vspace{5mm}
\end{abstract}

\section{Introduction}

\label{sec:intro}
\subsubsection*{Model reduction and hyper-reduction}
In spite of the ever increasing computational power and digital storage
capabilities, reduced-order models (ROM) of Finite Element (FE) discretized
systems are a must when complex simulations are needed to support
design and optimization activities. Current state-of-the-art ROM techniques
are based on a reduced basis approach, i.e., the solution is sought
on a \emph{linear}, low-dimensional subspace spanned by a few, carefully
selected modes. The resulting residual is then projected onto the
same or a different subspace of the same dimension to obtain a well-posed
reduced system. When the original system of equations--often referred
to as high-fidelity model (HFM)--features symmetric operators, the
residual is projected on the same basis used for the approximation
of the solution. This procedure is known as the Bubnov-Galerkin
projection. The dimensionality reduction is guaranteed by the limited
number of reduction variables as compared to the number of degrees
of freedom in the HFM.

Assuming granted accuracy, projection-based ROMs are not able to deliver significant speedups, when applied to nonlinear systems. This is due to the fact
that the computation of the reduced nonlinear terms at each solver
iteration scales with the dimensionality of the HFM rather than with that
of the reduction subspace. In the case of polynomial nonlinearities, one might consider pre-computing the reduced nonlinear term \cite{Barbic_2005,mignolet,MasterThesis}, leading to higher-order tensors. Typically, FE models may also contain rotational DOFs, where the associated nonlinearity cannot be pre-computed and stored in the form of tensors. Furthermore, for high polynomial degrees, the computations with higher-order tensors can become intractable, even more so than the HFM simulations \cite{MasterThesis}. For such cases, in the most general setting, the above-mentioned scalability issue is successfully alleviated by a range of \emph{hyper-reduction} techniques,
which aim at significantly reducing the cost of reduced nonlinear
term evaluation by computing them only at few, selected elements or
nodes of the FE models, and cheaply approximating the missing information. 

The Gappy POD method \cite{GappyPOD}, first introduced for image reconstruction and later applied in dynamical systems \cite{Carlberg_2010,Carlberg_2013}, was one of the first instances of hyper-reduction. However, as observed by Farhat et al. \cite{ECSW1}, the term hyper-reduction was introduced much later in \cite{Ryckelynck_2005}, where a selection of the integration point in a FE model is performed in an a priori manner. The empirical interpolation method (EIM) \cite{EIM} approximates a non-affine function using a set of basis functions over a continuous, bounded domain, and hence finds applications for reduced basis representations of parameterized partial differential equations (PDEs). The discrete empirical interpolation method (DEIM) \cite{DEIM} achieves this on a discrete state space. For FE applications, the unassembled DEIM or UDEIM \cite{UDEIM} was shown to be more efficient, where the unassembled element-level nonlinear force vector is used in the DEIM algorithm to return a small set of elements (instead of nodes) over which the nonlinearity is evaluated. For conservative/port-Hamiltonian systems, however, the DEIM (or UDEIM) leads to a loss of numerical stability during time integration. Though a symmetrized version of DEIM has been proposed recently \cite{symDEIM} to avoid this issue, its potential for FE-based applications remains unexplored. To this end, the energy-conserving sampling and weighting (ECSW) hyper-reduction method \cite{ECSW1} is remarkable. The ECSW method directly approximates the \emph{reduced} nonlinear term and is particularly well-suited for FE-based structural dynamics applications, as it preserves the symmetry of the operators involves and leads to numerical stability. 

The main idea of ECSW is simple yet powerful:
only few elements are interrogated to compute the nonlinear forces,
and weights (obtained using training) are attached to them to match
the virtual work done by their full counterparts onto virtual displacements
given by the linear reduction basis vectors. Albeit being efficient and relevant for FE-based mechanical systems, the ECSW hyper-reduction method
is proposed  in the context of \emph{linear}
mappings from the space of reduced variables. To the best of the authors' knowledge, this
is a common feature of all other hyper-reduction techniques in the literature, as well. 

\subsubsection*{Nonlinear manifolds in model reduction}
It is quite natural, however, for the dynamics of general nonlinear
systems to evolve on inherently nonlinear manifolds,
and not over linear subspaces (cf. \cite{slowfast} for a discussion). Often, a nonlinear manifold relevant for system dynamics may not be
embeddable in a low-dimensional linear subspace. It is then easy to
see that the linear-mapping-based ROM would either fail to capture
the system response, or require frequent basis-updates (online) to
account for the nonlinearity of the manifold, making the reduction process cumbersome and possibly redundant due to high computational costs in setting up such a ROM. Many methods allow for model reduction using attracting nonlinear manifolds in nonlinear dynamical systems -- from a theoretical, as well as a data-driven, statistical perspective. 

The approximate inertial manifold (AIM) provides an approximation of the solution
of a PDE over sufficiently
long time scales, using the theory of inertial manifolds. This is done by enslaving the amplitudes of fast modes in the system as a smooth graph over the amplitudes of a few selected modes while assuming that the corresponding inertial forces are not excited (see \cite{Ding_2012,Matthies2003}, for instance). The so called static condensation approach (cf. \cite{mignolet}) achieves the same objective without the projection of the governing equations on to  eigenfunctions. This is usually done for systems exhibiting a dichotomy of unknowns, such as the axial (in-plane) and transverse (out-of-plane) unknowns in a beam or a plate. From the perspective of nonlinear dynamics, a good overview of nonlinear model reduction techniques can be found in \cite{Rega_2005} and the references therein. 

Over the past decade or so, the use of nonlinear manifolds in model reduction has received a significant boost by many data-driven techniques, broadly referred to as \emph{nonlinear dimensionality reduction} (NLDR) or \emph{manifold learning}, see \cite{vdMaaten2009}, for instance. These techniques aim at constructing low dimensional manifolds from high-dimensional data snapshots. In a ROM perspective, these snapshots may be extracted from simulations
of the underlying HFM, and the chosen NLDR is deemed to provide a
nonlinear mapping with the required smoothness \cite{Millan2013}.
In essence, these methods provide a nonlinear generalization of the
Proper Orthogonal Decomposition (POD), a well-known technique which delivers
an optimal linear subspace for the dimensionality reduction. 

The use of nonlinear manifolds in model reduction is expected to not
only further reduce the dimensionality of the ROM in contrast to its linear
mapping counterpart, but also shed light on the underlying system behaviour.
This is especially true for the case of thin walled structures,
which exhibit a time scale separation between slow, out-of-plane;
and fast, in-plane dynamics. The fast dynamics in such systems can
often be enslaved statically to the slow one, in a nonlinear fashion
\cite{Jain2017}. Recent efforts \cite{qm,qm2} have shown that the geometrically
nonlinear dynamics of structures, characterized by bending-stretching
coupling (as for instance, thin walled components) can be effectively
captured using a quadratic manifold (QM), constructed using model
properties rather then HFM snapshots. 

\subsubsection*{Our contribution}
The nonlinear-manifold-based ROMs are usually obtained by projecting
the HFM on the tangent space of the manifold, in a virtual work sense.
This generates configuration-dependent inertial terms for the ROM in addition to the already existing projected nonlinear terms. For general nonlinear systems, the evaluation and projection of the
nonlinearity remains a bottleneck for computations, much like the
linear mapping counterpart. As discussed above, the state-of-the-art
hyper-reduction techniques allow for fast approximation of the (projected)
nonlinearity in the case of linear-mapping-based ROMs. This functionality, however, has been missing in the context of nonlinear-manifold-based ROMs. 

In this contribution, we generalize the application of the ECSW hyper-reduction
method to nonlinear-manifold-based ROMs. This work is essentially inspired from the ideas in the seminal work of Farhat et al. \cite{ECSW1,ECSW2}, where ECSW method was introduced in the context of linear mappings. Following similar lines as in \cite{ECSW2}, we further show that
our extension of the ECSW (hereafter referred to as \emph{extended} ECSW:
EECSW, for comparison purposes) preserves the Lagrangian structure associated with the Hamilton's
principle. A natural consequence of this property is numerical
stability during time integration, as shown in \cite{ECSW2}. The proposed method
is tested, first on an academically simple example, and then on a
realistic model of a wing using the recently proposed QM for reduction.
Our numerical results indicate that better speed ups can be achieved,
as compared to the ones obtained with a linear POD subspace reduction
equipped with ECSW. Interestingly, EECSW selects
fewer elements for the hyper-reduction as compared to its linear counterpart, and more than compensates the extra computational costs associated to the configuration dependent
inertial terms, which are not present in the case of linear subspace reduction.

This paper is organized as follows. The concept of projection-based
ROM on nonlinear mappings is briefly reviewed in Section 2. The extension
of ECSW for nonlinear mappings is the tackled in Section 3, where
also the preservation of the Lagrangian structure of the thus-obtained
ROM is shown. Numerical results showing the effectiveness of the method
are reported in Section 4, and finally, the conclusions are given
in Section 5.

\section{Model reduction using nonlinear mappings}

The partial differential equations (PDEs) for momentum balance in
a structural continuum are usually discretized along the spatial
dimensions using finite elements to obtain a system of second-order ordinary differential
equations (ODEs). Along with the initial conditions for generalized
displacements and velocities, these ODEs govern the response of the
underlying structure. More specifically, this response can be described
by the solution to an initial value problem (IVP) of the following
form:

\begin{gather}
\mathbf{\begin{gathered}\mathbf{M}\ddot{\mathbf{u}}(t)+\mathbf{C}\dot{\mathbf{u}}(t)-\mathbf{f}(\mathbf{u}(t))=\mathbf{f}^{ext}(t)\,,\\
	\mathbf{u}(t_{0})=\mathbf{u}_{0},\,\dot{\mathbf{u}}(t_{0})=\mathbf{v}_{0},
	\end{gathered}
}\label{eq:HFM}
\end{gather}
where the solution $\mathbf{u}(t)\in\mathbb{R}^{n}$ is a high-dimensional
generalized displacement vector, $\mathbf{M}\in\mathbb{R}^{n\times n}$
is the mass matrix; $\mathbf{C}\in\mathbb{R}^{n\times n}$ is the
damping matrix; $\mathbf{f}:\mathbb{R}^{n}\mapsto\mathbb{R}^{n}$
gives the nonlinear elastic internal force as a function of of the
displacement $\mathbf{u}$ of the structure; and $\mathbf{f}^{ext}(t)\in\mathbb{R}^{n}$
is the time dependent external load vector. The system (\ref{eq:HFM}),
referred to as the high-fidelity model (HFM), exhibits prohibitive
computational costs for large values of the dimension $n$ of the
system. The classical notion of model reduction aims to reduce this
dimensionality by introducing a linear mapping onto a suitable low-dimensional
invariant subspace. However, as discussed in the Introduction, more
recent advances in model reduction allow for the detection of system
dynamics evolving over nonlinear manifolds. The idea of a projection-based
reduced-order model using nonlinear mappings was presented in \cite{qm},
using a nonlinear mapping such as 
\begin{equation}
\mathbf{u}(t)\approx\boldsymbol{\Gamma}(\mathbf{q}(t)),\label{eq:NLM}
\end{equation}
where $\boldsymbol{\Gamma}:\mathbb{R}^{m}\mapsto\mathbb{R}^{n}$ is
the nonlinear mapping function. In general the mapping (\ref{eq:NLM}),
when substituted into the governing equations (\ref{eq:HFM}), results
in a non-zero residual $\mathbf{r}$ as
\[
\mathbf{M}\ddot{\boldsymbol{\Gamma}}\left(\mathbf{q}(t)\right)+\mathbf{C}\dot{\boldsymbol{\Gamma}}\left(\mathbf{q}(t)\right)-\mathbf{f}\left(\boldsymbol{\Gamma}\left(\mathbf{q}(t)\right)\right)-\mathbf{f}^{ext}(t)=\mathbf{\mathbf{r}}.
\]
The virtual work principle for a kinematically admissible displacement,
$\delta\mathbf{u}=\frac{\partial\boldsymbol{\Gamma}(\mathbf{q})}{\partial\mathbf{q}}\delta\mathbf{q}$,
over the nonlinear manifold leads to 
\[
\mathbf{\mathbf{r}}\cdot\delta\mathbf{u}=0
\]
\[
\implies\left[\mathbf{M}\ddot{\boldsymbol{\Gamma}}\left(\mathbf{q}(t)\right)+\mathbf{C}\dot{\boldsymbol{\Gamma}}\left(\mathbf{q}(t)\right)-\mathbf{f}\left(\boldsymbol{\Gamma}\left(\mathbf{q}(t)\right)\right)\right]\cdot\delta\mathbf{u}=\mathbf{f}^{ext}(t)\cdot\delta\mathbf{u}\,.
\]
Physically, this means that the residual force is orthogonal to the
tangent space of the manifold. This gives the corresponding reduced-order
model as 
\begin{equation}
\mathbf{P}_{\boldsymbol{\Gamma}}^{T}(\mathbf{q})\left[\mathbf{M}\ddot{\boldsymbol{\Gamma}}\left(\mathbf{q}(t)\right)+\mathbf{C}\dot{\boldsymbol{\Gamma}}\left(\mathbf{q}(t)\right) - \mathbf{f}\left(\boldsymbol{\Gamma}\left(\mathbf{q}(t)\right)\right)\right]=\mathbf{P}_{\boldsymbol{\Gamma}}^{T}(\mathbf{q})\mathbf{f}^{ext}(t),\label{eq:NLROM}
\end{equation}
where $\mathbf{P}_{\boldsymbol{\Gamma}}(\mathbf{q}):=\frac{\partial\boldsymbol{\Gamma}(\mathbf{q})}{\partial\mathbf{q}}\in\mathbb{R}^{n\times m}$
represents the tangent space of the nonlinear manifold in which the
system dynamics is expected to evolve. Hereafter, we shall omit the dependence of $ \mathbf{P}_{\boldsymbol{\Gamma}} $ on $ \mathbf{q} $. The acceleration and velocity
vectors over the manifold can be evaluated using the chain rule as
\begin{align*}
\dot{\boldsymbol{\Gamma}}\left(\mathbf{q}(t)\right) & =\frac{d}{dt}\boldsymbol{\Gamma}\left(\mathbf{q}(t)\right)=\frac{\partial\boldsymbol{\Gamma}(\mathbf{q})}{\partial\mathbf{q}}\cdot\dot{\mathbf{q}},\\
\ddot{\boldsymbol{\Gamma}}\left(\mathbf{q}(t)\right) & =\frac{d}{dt}\dot{\boldsymbol{\Gamma}}\left(\mathbf{q}(t)\right)=\frac{\partial\boldsymbol{\Gamma}(\mathbf{q})}{\partial\mathbf{q}}\cdot\ddot{\mathbf{q}}+\left[\frac{\partial^{2}\boldsymbol{\Gamma}(\mathbf{q})}{\partial\mathbf{q}\partial\mathbf{q}}\cdot\dot{\mathbf{q}}\right]\cdot\dot{\mathbf{q}}\,.
\end{align*}
Using the above relations, \eqref{eq:NLROM} can also be expressed as
\begin{equation}
\mathbf{M}_r(\mathbf{q})\,\ddot{\mathbf{q}} + \mathbf{C}_r(\mathbf{q},\dot{\mathbf{q}})\, \dot{\mathbf{q}} - \mathbf{f}_r(\mathbf{q}) = \mathbf{P}_{\boldsymbol{\Gamma}}^{T}\mathbf{f}^{ext}(t)\,,\label{eq:NLROM2} 
\end{equation}
where $ \mathbf{M}_r(\mathbf{q}) = \mathbf{P}_{\boldsymbol{\Gamma}}^{T}\mathbf{M}\mathbf{P}_{\boldsymbol{\Gamma}} $, $ \mathbf{C}_r(\mathbf{q},\dot{\mathbf{q}}) = \mathbf{P}_{\boldsymbol{\Gamma}}^T\mathbf{C}\mathbf{P}_{\boldsymbol{\Gamma}} +\mathbf{P}_{\boldsymbol{\Gamma}}^T \mathbf{M}\left( \frac{\partial^{2}\boldsymbol{\Gamma}(\mathbf{q})}{\partial\mathbf{q}\partial\mathbf{q}}\cdot\dot{\mathbf{q}}\right)  $ and $ \mathbf{f}_r(\mathbf{q}) = \mathbf{P}_{\boldsymbol{\Gamma}}^T\mathbf{f}(\boldsymbol{\Gamma}(\mathbf{q})) $.

Note that for $\boldsymbol{\Gamma}$ being a linear function, such
that $\boldsymbol{\Gamma}(\mathbf{q})=\mathbf{V}\mathbf{q}$ (for
some $\mathbf{V}\in\mathbb{R}^{n\times m}$ spanning a suitable low
dimensional subspace), the ROM in (\ref{eq:NLROM}) reduces to the
familiar Galerkin-projection-based ROMs as 
\begin{equation}
\mathbf{V}^{T}\mathbf{M}\mathbf{V}\ddot{\mathbf{q}}(t)+\mathbf{V}^{T}\mathbf{C}\mathbf{V}\dot{\mathbf{q}}(t)-\mathbf{V}^{T}\mathbf{f}(\mathbf{Vq}(t))=\mathbf{V}^{T}\mathbf{f}^{ext}(t).\label{eq:GalerkinROM}
\end{equation}
During time integration--usually referred to as the \emph{online phase}--of (\ref{eq:GalerkinROM}), the computational cost associated to
the evaluation of the linear terms in (\ref{eq:GalerkinROM}) scales
only with the number of reduced variables $m$, since the corresponding
reduced operators can be pre-computed \emph{offline}. However, this
is not the case for the computation of the reduced nonlinear term
$\mathbf{V}^{T}\mathbf{f}(\mathbf{Vq}(t))$. Specifically, for FE-based
applications, this evaluation is usually carried out \emph{online,}
in the following manner: 
\begin{equation}
\hat{\mathbf{f}}(\mathbf{q}):=\mathbf{V}^{T}\mathbf{f}(\mathbf{Vq})=\sum_{e=1}^{n_{e}}\mathbf{V}_{e}^{T}\mathbf{f}_{e}(\mathbf{V}_{e}\mathbf{q}),\label{eq:felsum}
\end{equation}
where $\mathbf{f}_{e}(\mathbf{u_{e}})\in\mathbb{R}^{N_{e}}$ is the
contribution of the element $e$ towards the vector $\mathbf{f}(\mathbf{u})$
($N_{e}$ being the number of DOFs for the element $e$), $\mathbf{V}_{e}$
is the restriction of $\mathbf{V}$ to the rows indexed by the DOFs
corresponding to $e$, and $n_{e}$ is the total number of elements
in the structure. It is then easy to see from (\ref{eq:felsum}),
that the cost associated to the computation of the reduced nonlinear
force $\hat{\mathbf{f}}(\mathbf{q})$ scales linearly with the total
number of elements $n_{e}$ in the structure, which is potentially
high for large systems. Thus, despite the reduction in dimensionality
achieved in (\ref{eq:GalerkinROM}), the evaluation of the reduced
nonlinear term $\hat{\mathbf{f}}(\mathbf{q})$ emerges as a new bottleneck
for the fast prediction of system response using the ROM (\ref{eq:GalerkinROM}).
Similar conclusions regarding computational bottlenecks hold for the
nonlinear-reduction-based ROM (\ref{eq:NLROM}), as well.

\section{Hyper-reduction and Extended ECSW}

Hyper-reduction techniques help mitigate the above-stated high computational
costs by approximation of the reduced-nonlinear term in a computationally
affordable manner. As discussed in the Introduction, the ECSW \cite{ECSW1} is a remarkable
method for hyper-reduction of linear-mapping-based ROMs, such as that in (\ref{eq:GalerkinROM}). After briefly summarizing ECSW, we propose here a natural extension to ECSW, which facilitates hyper-reduction of nonlinear-manifold-based ROM, such as in (\ref{eq:NLROM}). Much
like the ECSW, we further show in Section \ref{sec:Lag_struct} that the proposed EECSW
preserves the Lagrangian structure associated to the underlying mechanical system and, hence, leads to numerical stablity in time integration.

In ECSW \cite{ECSW1}, the computationally expensive reduced nonlinear forces such as $ \hat{\mathbf{f}}(\mathbf{q}) $ (cf. \eqref{eq:felsum}) in the Galerkin-projection-based ROMs, e.g., \eqref{eq:GalerkinROM} is approximated as 
\begin{equation}
\text{(ECSW)}\quad\hat{\mathbf{f}}(\mathbf{q})\approx\tilde{\mathbf{f}}(\mathbf{q}):=\sum_{e\in E}\xi_{e}\mathbf{V}_{e}^{T}\mathbf{f}_{e}(\mathbf{V}_{e}\mathbf{q}),\label{eq:ECSWapprox}
\end{equation}
where $\xi_{e}\in\mathbb{R}^{+}$ is a positive weight given to every
element $e\in E$. Clearly, the evaluation of $\tilde{\mathbf{f}}(\mathbf{q})$
in (\ref{eq:ECSWapprox}) comes at a fraction of the computational
cost associated to (\ref{eq:felsum}), for $|E|\ll n_{e}$. This is made possible with weights which are chosen using \emph{training} to ensure a good approximation of the original sum over all $n_{e}$ elements, while keeping the element sample small using sparse algorithms in literature. The element-set $ E $ is sometimes referred to as the \emph{reduced mesh} due to its low cardinality relative to the total number of elements $ n_e $. Farhat et al. \cite{ECSW2} further show that hyper-reduction done in this manner preserves Lagrangian structure of the linear Galerkin-projection-based ROMs. 

We observe that analogous to (\ref{eq:felsum}), the projected nonlinear force (possibly a combination of inertial, damping or elastic forces) in the nonlinear ROM (\ref{eq:NLROM}) can be expressed as a summation of the element-level contributions as
\begin{equation}
\boldsymbol{\mathfrak{h}}_r(\mathbf{q},\dot{\mathbf{q}},\ddot{\mathbf{q}}):=\mathbf{P}_{\boldsymbol{\Gamma}}^{T}\boldsymbol{\mathfrak{h}}\left(\boldsymbol{\Gamma}(\mathbf{q}),\dot{\boldsymbol{\Gamma}}\left(\mathbf{q}\right),\ddot{\boldsymbol{\Gamma}}\left(\mathbf{q}\right)\right)=\sum_{e=1}^{n_{e}}\left(\mathbf{P}_{\boldsymbol{\Gamma}}(\mathbf{q})\right)_{e}^{T}\boldsymbol{\mathfrak{h}}_{e}\left(\boldsymbol{\Gamma}_{e}(\mathbf{q}),\dot{\boldsymbol{\Gamma}}_{e}\left(\mathbf{q}\right),\ddot{\boldsymbol{\Gamma}}_{e}\left(\mathbf{q}\right)\right),\label{eq:nlfsum}
\end{equation}
where $\boldsymbol{\mathfrak{h}}_r\left(\mathbf{q},\dot{\mathbf{q}},\ddot{\mathbf{q}}\right)$
is the part of the ROM (\ref{eq:NLROM}) which is potentially unaffordable to
compute (cost scales with $n$ or $n_{e}$), $  \boldsymbol{\mathfrak{h}} $ is its full counterpart,   with the  $  \boldsymbol{\mathfrak{h}}_e $
having the usual meaning for the element-level contribution from the element $e$. Just like in ECSW, the summation in (\ref{eq:nlfsum}) may be approximated as 
\begin{equation}
\text{(EECSW)}\quad\boldsymbol{\mathfrak{h}}_r(\mathbf{q},\dot{\mathbf{q}},\ddot{\mathbf{q}})\approx\tilde{\boldsymbol{\mathfrak{h}}}_r(\mathbf{q},\dot{\mathbf{q}},\ddot{\mathbf{q}})=\sum_{e\in E}\xi_{e}\left(\mathbf{P}_{\boldsymbol{\Gamma}}(\mathbf{q})\right)_{e}^{T}\boldsymbol{\mathfrak{h}}_{e}\left(\boldsymbol{\Gamma}_{e}(\mathbf{q}),\dot{\boldsymbol{\Gamma}}_{e}\left(\mathbf{q}\right),\ddot{\boldsymbol{\Gamma}}_{e}\left(\mathbf{q}\right)\right),\label{eq:EECSWaprox}
\end{equation}
where $\xi_{e}\in\mathbb{R}^{+}$ is again a positive weight given to every
element $e$ in the reduced mesh $ E $. The element sampling and weighing are discussed in Section \ref{sec:elSW}

One easily observable distinction of EECSW from its linear mapping counterpart is that in ECSW, it is possible to precompute the reduced linear operators in the hyper-reduced model, such as the reduced mass, damping and stiffness matrices $ \mathbf{V}^T\mathbf{M}\mathbf{V} $, $ \mathbf{V}^T\mathbf{C}\mathbf{V} $ and $ \mathbf{V}^T\mathbf{K}\mathbf{V} $, respectively. Due to the nonlinear mapping and projection on the tangent space, we see that EECSW doesn't
feature reduced-linear operators, even for the linear mass, stiffness
and damping matrices in the HFM. Thus, in principle, one could include
all the terms on the left hand of (\ref{eq:NLROM}) in $ \boldsymbol{\mathfrak{h}}_r $, i.e.,
\[
\boldsymbol{\mathfrak{h}}_r(\mathbf{q},\dot{\mathbf{q}},\ddot{\mathbf{q}})=\mathbf{P}_{\boldsymbol{\Gamma}}^{T}\left[\mathbf{M}\ddot{\boldsymbol{\Gamma}}\left(\mathbf{q}(t)\right)+\mathbf{C}\dot{\boldsymbol{\Gamma}}\left(\mathbf{q}(t)\right)-\mathbf{f}\left(\boldsymbol{\Gamma}\left(\mathbf{q}(t)\right)\right)\right].
\]

It is easy to see that this idea is a straight-forward generalization
of ECSW to nonlinear-manifold-based ROMs with projection to the tangent
space, as given in (\ref{eq:NLROM}). Indeed, for $\boldsymbol{\Gamma}(\mathbf{q})=\mathbf{V}\mathbf{q}$,
hyper-reduction of ROM (\ref{eq:GalerkinROM}) results in the same approximation
of the projected internal force as in \eqref{eq:felsum}.

\subsection{Element sampling and weight selection}
\label{sec:elSW}
The elements and weights are determined to approximate virtual work
over chosen training sets which generally come from full solution
run(s). We assume that $\left\{ \mathbf{u}^{(i)},\dot{\mathbf{u}}^{(i)},\ddot{\mathbf{u}}^{(i)}\right\} ,\,i\in\{1,\dots,n_{t}\}$
represents a set of available training vectors. If the solution to
the HFM, indeed, evolves over a nonlinear manifold as given by (\ref{eq:NLM}),
then we must be able to find reduced unknowns $\left\{ \mathbf{q}^{(i)},\dot{\mathbf{q}}^{(i)},\ddot{\mathbf{q}}^{(i)}\right\} $ 
such that 
\begin{equation}
\left.\begin{aligned}\mathbf{u}^{(i)} & =\boldsymbol{\Gamma}\left(\mathbf{q}^{(i)}\right)\\
\dot{\mathbf{u}}^{(i)} & =\mathbf{P}_{\boldsymbol{\Gamma}}\left(\mathbf{q}^{(i)}\right)\dot{\mathbf{q}}^{(i)}\\
\ddot{\mathbf{u}}^{(i)} & =\mathbf{P}_{\boldsymbol{\Gamma}}^{T}\left(\mathbf{q}^{(i)}\right)\ddot{\mathbf{q}}^{(i)}+\left[\left.\frac{\partial^{2}\boldsymbol{\Gamma}(\mathbf{q})}{\partial\mathbf{q}\partial\mathbf{q}}\right|_{\mathbf{q}=\mathbf{q}^{(i)}}\cdot\dot{\mathbf{q}}^{(i)}\right]\cdot\dot{\mathbf{q}}^{(i)}
\end{aligned}\right\rbrace
\quad\forall i\in\{1,\dots,n_{t}\}\,.\label{eq:train}
\end{equation}
As will be clear later in this section, the optimal elements and weights
corresponding to a reduced mesh needs the set of reduced vectors $\left\{ \mathbf{q}^{(i)},\dot{\mathbf{q}}^{(i)},\ddot{\mathbf{q}}^{(i)}\right\} $
during the training stage. In case these are available from a previous
ROM simulation, these would be directly useful. Furthermore, if the nonlinear manifold is obtained from solution snapshots using manifold-learning procedures such as the local maximum entropy approximants, as done in, e.g, \cite{Millan2013}, then the reduced vectors would be directly available from the manifold-learning procedure. However, if the manifold is obtained from simulation-free approaches (cf. \cite{qm,qm2}), then a nonlinear system of algebraic equations (\ref{eq:train}) can be solved for the reduced training vectors $\mathbf{q}^{(i)},\dot{\mathbf{q}}^{(i)},\text{and }\ddot{\mathbf{q}}^{(i)}$. This system is under-constrained
($m\ll n$) and , hence, the sought $\mathbf{q}^{(i)}$ can
be obtained from the solution to a nonlinear least-squares procedures, e.g., the gradient descent method, the Gauss-Newton methods, the Levenberg-Marquardt method (which is also employed in the standard \texttt{LSQNONLIN} routine in MATLAB); we refer the reader to \cite{lines84} for a review of standard methods for nonlinear least squares problems.

Geometrically, the first equation in the system (\ref{eq:train}) gives the location of the data point on the manifold, whereas as the second and the third equations give the trajectory information (tangent and curvature along the trajectory). Thus, the reduced coordinates of the point on the manifold can be identified independent of the velocity and acceleration information by solving the equation
\begin{equation}
\mathbf{q}^{(i)}=\arg\min_{\mathbf{q}\in\mathbb{R}^{m}}\left\Vert \boldsymbol{\Gamma}\left(\mathbf{q}\right)-\mathbf{u}^{(i)}\right\Vert _{2}^{2}.\label{eq:LSQ}
\end{equation}
Though the nonlinear least squares problem \eqref{eq:LSQ} would add to the offline costs, note that this procedure is embarassingly parallelizable for each training snapshot. Once the reduced coordinates $\mathbf{q}^{(i)}$ are obtained on the manifold, the second equation in (\ref{eq:train}) can be solved in a least squares sense for $\dot{\mathbf{q}}^{(i)}$
as 
\begin{equation}
\label{eq:train_vel}
\dot{\mathbf{q}}^{(i)}=\left[\mathbf{P}_{\boldsymbol{\Gamma}}^{T}\left(\mathbf{q}^{(i)}\right)\mathbf{P}_{\boldsymbol{\Gamma}}\left(\mathbf{q}^{(i)}\right)\right]^{-1}\mathbf{P}_{\boldsymbol{\Gamma}}^{T}\left(\mathbf{q}^{(i)}\right)\dot{\mathbf{u}}^{(i)}.
\end{equation}
Finally, we can obtain the corresponding reduced accelerations as
\begin{equation}
\label{eq:train_acc}
\ddot{\mathbf{q}}^{(i)}=\left[\mathbf{P}_{\boldsymbol{\Gamma}}^{T}\left(\mathbf{q}^{(i)}\right)\mathbf{P}_{\boldsymbol{\Gamma}}\left(\mathbf{q}^{(i)}\right)\right]^{-1}\mathbf{P}_{\boldsymbol{\Gamma}}^{T}\left(\mathbf{q}^{(i)}\right)\left(\ddot{\mathbf{u}}^{(i)}-\left[\left.\frac{\partial^{2}\boldsymbol{\Gamma}(\mathbf{q})}{\partial\mathbf{q}\partial\mathbf{q}}\right|_{\mathbf{q}=\mathbf{q}^{(i)}}\cdot\dot{\mathbf{q}}^{(i)}\right]\cdot\dot{\mathbf{q}}^{(i)}\right).
\end{equation}
Alternatively, one might consider solving a single least squares problems involving all the reduced vectors for displacements, velocities and accelerations, resulting in a system which is three times in dimensionality. However, the sequential procedure proposed in \eqref{eq:LSQ}-\eqref{eq:train_acc} is computationally more effective. Note that for $\boldsymbol{\Gamma}$ being a linear function, such
as $\boldsymbol{\Gamma}(\mathbf{q})=\mathbf{V}\mathbf{q}$, the system
(\ref{eq:LSQ}) reduces to the familiar least squares solution of
(\ref{eq:train}), available in closed form as 
\begin{align*}
\mathbf{q}^{(i)} & =(\mathbf{V}^{T}\mathbf{V})^{-1}\mathbf{V}^{T}\mathbf{u}^{(i)},\\
\dot{\mathbf{q}}^{(i)} & =(\mathbf{V}^{T}\mathbf{V})^{-1}\mathbf{V}^{T}\dot{\mathbf{u}}^{(i)},\\
\ddot{\mathbf{q}}^{(i)} & =(\mathbf{V}^{T}\mathbf{V})^{-1}\mathbf{V}^{T}\ddot{\mathbf{u}}^{(i)}.
\end{align*}
Once the reduced training vectors  $\left\{ \mathbf{q}^{(i)},\dot{\mathbf{q}}^{(i)},\ddot{\mathbf{q}}^{(i)}\right\} $ are obtained,
the element level contribution of projected internal force for each
of the training vectors can be assembled in a matrix $\mathbf{G}$
as follows: 
\begin{equation}
\begin{gathered}\mathbf{G}=\begin{bmatrix}\mathbf{g}_{11} & \dots & \mathbf{g}_{1n_{e}}\\
\vdots & \ddots & \vdots\\
\mathbf{g}_{n_{t}1} & \dots & \mathbf{g}_{n_{t}n_{e}}
\end{bmatrix}\in\mathbb{R}^{mn_{t}\times n_{e}},\hspace{1cm}\mathbf{b}=\begin{bmatrix}\mathbf{b}_{1}\\
\vdots\\
\mathbf{b}_{n_{t}}
\end{bmatrix}\in\mathbb{R}^{mn_{t}}\,,\\
\mathbf{g}_{ie}=\left(\mathbf{P}_{\boldsymbol{\Gamma}}\left(\mathbf{q}^{(i)}\right)\right)_{e}^{T}\boldsymbol{\mathfrak{h}}_{e}\left(\boldsymbol{\Gamma}_{e}(\mathbf{q}^{(i)}),\dot{\boldsymbol{\Gamma}}_{e}\left(\mathbf{q}^{(i)}\right),\ddot{\boldsymbol{\Gamma}}_{e}\left(\mathbf{q}^{(i)}\right)\right),\qquad\mathbf{b}_{i}=\boldsymbol{\mathfrak{h}}_{r}\left(\mathbf{q}^{(i)}\right)=\sum_{e=1}^{n_{e}}\mathbf{g}_{ie}\,\\
\quad\forall i\in\{1,\dots,n_{t}\},\quad e\in\{1,\dots,n_{e}\}\,.
\end{gathered}
\label{eq:Gbmatrix}
\end{equation}
Analogous to ECSW, the set of elements and weights is then obtained
by a sparse solution to the following non-negative least-squares (NNLS)
problem 
\begin{equation}
(P1):\boldsymbol{\zeta}=\text{arg}\min_{\tilde{\boldsymbol{\zeta}}\in\mathbb{R}^{n_{e}},\tilde{\boldsymbol{\zeta}}\ge\mathbf{0}}\|\mathbf{G}\tilde{\boldsymbol{\zeta}}-\mathbf{b}\|_{2}^{2},\label{eq:NNLS}
\end{equation}
A sparse solution to $(P1)$ returns a sparse vector $\boldsymbol{\zeta}$,
the non-zero entries of which form the reduced mesh $E$ used in (\ref{eq:ECSWapprox})
as 
\[
E=\{e:\zeta_{e}>0\}.
\]
An optimally sparse solution to $(P1)$ is NP-hard to obtain. However,
a greedy-approach-based algorithm \cite{sNNLS}, which finds a sub-optimal
solution, has has been found to deliver an effective reduced mesh
$E$ \cite{ECSW1}. We have included
its pseudo-code in Algorithm \ref{alg:sNNLS}, where $\boldsymbol{\zeta}_E $ and $ \mathbf{G}_E $ denote, respectively, the restriction of $ \boldsymbol{\zeta} \in \mathbb{R}^{n_e} $ and column-wise restriction of $ \mathbf{G} $ to the elements in the \textit{active} subset $ E $. The set $ Z $ is the disjoint \textit{inactive} subset which contains the zero entry-indices of $ \boldsymbol{\xi} $ and $ \boldsymbol{\zeta} $. More recent work \cite{Chapman_2017} proposes and discusses different alternatives to accelerate the sampling procedure during hyper-reduction, including its parallel implementation to reduce offline costs. 
\begin{algorithm}[H]
	\begin{algorithmic}[1]
		
		\Require{$ \{\mathbf{u}^{(i)}\},\,i\in\{1,\dots,n_{t}\}, \tau $ (sparse NNLS tolerance)}
		
		\Ensure{$ \boldsymbol{\xi} \in \mathbb{R}^{n_e} $ sparse, $ E \subset \{1,\dots,n_e\}$} 	
		
		\Statex
		
		\For{$i \gets 1 \textrm{ to } n_t$} 
		
		\State $\mathbf{q}^{(i)}\gets\texttt{LSQNLIN}\left(\mathbf{u}^{(i)}\right)$ \Comment 
		$\texttt{LSQNLIN}$: solves the nonlinear least squares problem (\ref{eq:LSQ}),
		e.g., using gradient descent method (\ref{eq:grad_des})
		
		\EndFor
		
		\State $[\mathbf{G},\mathbf{b}]\gets\texttt{Construct\_G\_b}\left(\{\mathbf{q}^{(i)}\}_{1}^{n_{t}}\right)$ \Comment 
		$\texttt{Construct\_G\_b}$: assembles $\mathbf{G},\mathbf{b}$, as
		given by (\ref{eq:Gbmatrix})
		
		\State $\ensuremath{E\gets\emptyset},\ensuremath{Z\gets\{1,\dots,n_{e}\}},\ensuremath{\boldsymbol{\xi}\gets\mathbf{0}\in\mathbb{R}^{n_{e}}}$
		
		\While {$ \|\mathbf{G}\bs{\xi} - \mathbf{b}\|_2 > \tau \|\mathbf{b}\|_2$ } 	 \Comment sparse
		NNLS solver
		
		\State$\ensuremath{\boldsymbol{\mu}\gets\mathbf{G}^{T}(\mathbf{b}-\mathbf{G}\boldsymbol{\xi})}$
		
		\State$\ensuremath{[|\nu|,e]\gets\texttt{max}\left(\boldsymbol{\mu}\right)}$ \Comment $\texttt{max}$:
		returns maximum value in a vector followed by its location (index)
		
		\State$\ensuremath{E\gets E\cup\{e\}},\ensuremath{Z\gets Z\backslash\{e\}}$
		
		\While{\textbf{true}}
		
		\State$\boldsymbol{\zeta}_{E}\gets\mathbf{G}_{E}^{\dagger}\mathbf{b}$	\Comment  $\dagger$
		represents pseudo-inverse
		
		\State$\ensuremath{\boldsymbol{\zeta}_{Z}\gets\mathbf{0}}$
		
		\If {$ \bs{\zeta}_E > \mathbf{0} $} 
		
		\State$\ensuremath{\boldsymbol{\xi}\gets\boldsymbol{\zeta}}$
		
		\State$\textbf{break}$
		
		\EndIf
		
		\State$\ensuremath{\eta\gets\min\limits _{k\in E}\xi_{k}/(\xi_{k}-\zeta_{k})}$
		
		\State$\ensuremath{\boldsymbol{\xi}\gets\boldsymbol{\xi}+\eta(\boldsymbol{\zeta}-\boldsymbol{\xi})}$
		
		\State$\ensuremath{Z\gets\{i|\xi_{i}=0\}}$
		
		\State$\ensuremath{E\gets\{1,\dots,n_{e}\}\backslash Z}$
		
		\EndWhile
		
		\EndWhile
		
	\end{algorithmic}
	
	\caption{\label{alg:sNNLS}Training for EECSW }
\end{algorithm}

\subsection{Preservation of Lagrangian structure and numerical stability}
\label{sec:Lag_struct}
\subsubsection{Conservative systems}
Following similar steps as in \cite{ECSW2}, we show here that the
hyper-reduced model preserves the Lagrangian structure, leading to
numerical stability of the hyper-reduced model. For this purpose,
we restrict our discussion to a conservative system, where the Hamiltonian
principle represents the conservation of total mechanical energy as
follows: 
\begin{equation}
\mathcal{E}=\mathcal{T}(\dot{\mathbf{u}})+\mathcal{V}(\mathbf{u})=constant,
\end{equation}
where $\mathcal{T}$ and $\mathcal{V}$ are the functionals associated
to the kinetic energy and the potential energy of the system, respectively.
The kinetic energy is given by 
\begin{equation}
\mathcal{T}(\dot{\mathbf{u}})=\frac{1}{2}\dot{\mathbf{u}}^{T}\mathbf{M}\dot{\mathbf{u}} = \sum_{e=1}^{n_e}\underbrace{\frac{1}{2}\dot{\mathbf{u}}_e\mathbf{M}_e\dot{\mathbf{u}}_e}_{\mathcal{T}_e(\dot{\mathbf{u}}_e)},
\end{equation}
where $ \mathbf{M}_e \in \mathbb{R}^{N_e\times N_e}$ is the contribution of the element $ e $ towards the mass matrix and $ \mathcal{T}_e $ is the corresponding contribution towards the total kinetic energy. Furthermore, the potential energy associated to the elastic internal force $ \mathbf{f} $ and the conservative external force $ \mathbf{f}^{ext}_c $ can be written as 
\begin{gather}
\mathcal{V}(\mathbf{u})=\underbrace{\int-\mathbf{f}(\mathbf{u})\cdot\mathrm{d}\mathbf{u}}_{\mathcal{V}^{int}(\mathbf{u})}+\underbrace{\int \mathbf{f}^{ext}_c\cdot\mathrm{d}\mathbf{u}}_{\mathcal{V}^{ext}(\mathbf{u})}\,,
\end{gather}
such that
\begin{gather}
\mathbf{f}(\mathbf{u})=-\frac{\partial\mathcal{V}^{int}(\mathbf{u})}{\partial\mathbf{u}},\\
-\mathbf{f}^{ext}_c(\mathbf{u}) = -\frac{\partial\mathcal{V}^{ext}(\mathbf{u})}{\partial\mathbf{u}},
\end{gather}
and $\mathcal{V}^{int}$ and $\mathcal{V}^{ext}$ are the potential
energies associated to internal and external forces, respectively.
The conservative counterpart of the equations of motion (\ref{eq:HFM})
can then be expressed using Lagrange equations as: 
\begin{equation}
\frac{d}{dt}\left(\frac{\partial\mathcal{T}(\dot{\mathbf{u}})}{\partial \dot{\mathbf{u}}}\right) + \frac{\partial\mathcal{V}^{int}(\mathbf{u})}{\partial\mathbf{u}}=\mathbf{f}^{ext}_c\,.
\end{equation}
As shown in \eqref{eq:NLROM}, the reduced-order model, obtained by
introduction of a nonlinear mapping $\boldsymbol{\Gamma}$ and projection
of the resulting equations to the tangent space $\mathbf{P}_{\boldsymbol{\Gamma}}$
of the manifold, is given by 
\begin{gather}
\mathbf{P}_{\boldsymbol{\Gamma}}^{T}\left[\frac{d}{dt}\left(\frac{\partial\mathcal{T}(\dot{\boldsymbol{\Gamma}}(\mathbf{q}))}{\partial \dot{\boldsymbol{\Gamma}}(\mathbf{q})}\right) +
\frac{\partial\mathcal{V}^{int}(\boldsymbol{\Gamma}(\mathbf{q}))}{\partial\boldsymbol{\Gamma}(\mathbf{q})}\right]=\mathbf{P}_{\boldsymbol{\Gamma}}^{T}\mathbf{f}^{ext}_c,\label{eq:lag_cons_rom}
\end{gather}
where $\boldsymbol{\Gamma}:\mathbb{R}^{n}\mapsto\mathbb{R}^{m}$ describes
a general nonlinear mapping which approximates the solution $\mathbf{u}$
using a reduced set of unknowns $\mathbf{q}$. It is easy to see that \eqref{eq:lag_cons_rom}  is equivalent to ROM \eqref{eq:NLROM} in the conservative setting, i.e., 
\begin{gather}
\mathbf{P}_{\boldsymbol{\Gamma}}^{T}\left[\mathbf{M}\ddot{\boldsymbol{\Gamma}}(\mathbf{q}) +
\frac{\partial\mathcal{V}^{int}(\boldsymbol{\Gamma}(\mathbf{q}))}{\partial\boldsymbol{\Gamma}(\mathbf{q})}\right]=\mathbf{P}_{\boldsymbol{\Gamma}}^{T}\mathbf{f}^{ext}_c\label{eqn:QMred}\,.
\end{gather}
Recall that the ROM
obtained using a linear Galerkin projection is a special case of the general
setting in (\ref{eqn:QMred}) when a linear mapping $\boldsymbol{\Gamma}(\mathbf{q})=\mathbf{Vq}$
is used. It is well known that such
a linear Galerkin projection is structure-preserving (cf., e.g., \cite{lall_2003,ECSW2}). We now show that this
is also true for ROM (\ref{eqn:QMred}), obtained by projection on
to the tangent space of general nonlinear manifolds.

We can define the reduced potential and kinetic energy associated to the reduced system \eqref{eqn:QMred} as 
\begin{align}
\mathcal{V}_{r}(\mathbf{q}) &=\mathcal{V}\left(\boldsymbol{\Gamma}(\mathbf{q})\right)\,,\\
\mathcal{T}_{r}(\dot{\mathbf{q}},\mathbf{q})&=\mathcal{T}\left(\dot{\boldsymbol{\Gamma}}(\mathbf{q})\right)\,.
\end{align}
Then, the Hamiltonian associated to ROM \eqref{eqn:QMred} is given by 
\begin{align}
\mathcal{E}_{r}(\mathbf{q},\dot{\mathbf{q}}) & =\mathcal{V}_{r}(\mathbf{q}) + \mathcal{T}_{r}(\mathbf{q}),\nonumber \\
\implies\dot{\mathcal{E}}_{r}(\mathbf{q},\dot{\mathbf{q}},\ddot{\mathbf{q}}) & =\dot{\mathcal{T}}(\dot{\boldsymbol{\Gamma}}(\mathbf{q}))+\dot{\mathcal{V}}(\boldsymbol{\Gamma}(\mathbf{q}))\nonumber \\
& =\frac{\partial\mathcal{T}(\dot{\boldsymbol{\Gamma}}(\mathbf{q}))}{\partial\dot{\boldsymbol{\Gamma}}(\mathbf{q})}\ddot{\boldsymbol{\Gamma}}(\mathbf{q})+\frac{\partial\mathcal{V}^{int}(\boldsymbol{\Gamma}(\mathbf{q}))}{\partial\boldsymbol{\Gamma}(\mathbf{q})}\dot{\boldsymbol{\Gamma}}(\mathbf{q}) + \frac{\partial\mathcal{V}^{ext}(\boldsymbol{\Gamma}(\mathbf{q}))}{\partial\boldsymbol{\Gamma}(\mathbf{q})}\dot{\boldsymbol{\Gamma}}(\mathbf{q})\nonumber \\
& =(\dot{\boldsymbol{\Gamma}}(\mathbf{q}))^{T}\mathbf{M}\ddot{\boldsymbol{\Gamma}}(\mathbf{q})+(\dot{\boldsymbol{\Gamma}}(\mathbf{q}))^{T}\frac{\partial\mathcal{V}^{int}(\boldsymbol{\Gamma}(\mathbf{q}))}{\partial\boldsymbol{\Gamma}(\mathbf{q})}-(\dot{\boldsymbol{\Gamma}}(\mathbf{q}))^{T}\mathbf{f}^{ext}_c\nonumber \\
& =\dot{\mathbf{q}}^{T}\left(\mathbf{P}_{\boldsymbol{\Gamma}}^{T}\left[\mathbf{M}\ddot{\boldsymbol{\Gamma}}(\mathbf{q})+\frac{\partial\mathcal{V}^{int}(\boldsymbol{\Gamma}(\mathbf{q}))}{\partial\boldsymbol{\Gamma}(\mathbf{q})}-\mathbf{f}^{ext}_c\right]\right)\qquad(\because\dot{\boldsymbol{\Gamma}}(\mathbf{q})=\mathbf{P}_{\boldsymbol{\Gamma}}\dot{\mathbf{q}})\nonumber \\
& =0\qquad(\text{Using (\ref{eqn:QMred})})
\end{align}
Thus, $\mathcal{E}_{r}=constant$, and the total mechanical energy
is conserved for the ROM in (\ref{eqn:QMred}). It is also easy to see that the reduced equations can be directly obtained using the reduced Lagrangian $ \mathcal{L}_r := \mathcal{T}_r - \mathcal{V}_r$. This proves that the projection of the equations to the tangent
space of the manifold leads to preservation of this Lagrangian structure
associated to the corresponding Hamiltonian principle. Note that the projection of the full system on to any subspace other than the tangent space of the manifold would not lead to these desirable properties, in general.

In the hyper-reduced model, the reduced internal force from \eqref{eq:lag_cons_rom} is approximated using EECSW (cf. (\ref{eq:EECSWaprox})) as 
\begin{gather}
\label{eq:Vred}
\mathbf{P}_{\boldsymbol{\Gamma}}^{T}\frac{\partial\mathcal{V}^{int}(\boldsymbol{\Gamma}(\mathbf{q}))}{\partial\boldsymbol{\Gamma}(\mathbf{q})}\approx \frac{\partial\mathcal{\tilde{V}}_{r}^{int}(\mathbf{q})}{\partial\mathbf{q}} = \sum_{e\in E}\xi_{e}\left(\mathbf{P}_{\boldsymbol{\Gamma}}^{T}\right)_{e}\frac{\partial\mathcal{V}_{e}^{int}(\boldsymbol{\Gamma}_{e}(\mathbf{q}))}{\partial\boldsymbol{\Gamma}_{e}(\mathbf{q})}\,,
\end{gather}
where
\begin{equation}
\mathcal{\tilde{V}}_{r}^{int}(\mathbf{q})=\sum_{e\in E}\xi_{e}\mathcal{V}_{e}^{int}(\boldsymbol{\Gamma}_{e}(\mathbf{q}))\,
\end{equation}
can be identified as the potential energy associated to the hyper-reduced internal force. Furthermore, the inertial force can also be approximated using EECSW as
\begin{equation}
\label{eq:Tred}
\mathbf{P}_{\boldsymbol{\Gamma}}^{T}\frac{d}{dt}\left(\frac{\partial\mathcal{T}(\dot{\boldsymbol{\Gamma}}(\mathbf{q}))}{\partial \dot{\boldsymbol{\Gamma}}(\mathbf{q})}\right) \approx \frac{d}{dt}\left(\frac{\partial\mathcal{\tilde{T}}_{r}(\dot{\mathbf{q}},\mathbf{q})}{\partial \dot{\mathbf{q}}}\right)  =  
\sum_{e\in E} \xi_e \left(\mathbf{P}_{\boldsymbol{\Gamma}}^{T}\right)_{e} \frac{d}{dt} \frac{\partial\mathcal{T}_{e}(\dot{\boldsymbol{\Gamma}}_{e}(\mathbf{q}))}{\partial\dot{\boldsymbol{\Gamma}}_{e}(\mathbf{q})}=\sum_{e\in E} \xi_e \left(\mathbf{P}_{\boldsymbol{\Gamma}}^{T}\right)_{e} \mathbf{M}_e \ddot{\boldsymbol{\Gamma}}_e (\mathbf{q})\,,
\end{equation}
where
\begin{equation}
\mathcal{\tilde{T}}_{r}(\dot{\mathbf{q}},\mathbf{q})=\sum_{e\in E} \xi_{e} \mathcal{T}_{e}(\dot{\boldsymbol{\Gamma}}_{e}(\mathbf{q}))
\end{equation}
can be interpreted as the kinetic energy associated to the reduced mesh. 
To summarise, the hyper-reduced equations, using EECSW, for the ROM \eqref{eq:lag_cons_rom} are given as 
\begin{gather}
\sum_{e\in E} \xi_e \left(\mathbf{P}_{\boldsymbol{\Gamma}}^{T}\right)_{e} \mathbf{M}_e \ddot{\boldsymbol{\Gamma}}_e (\mathbf{q}) + \sum_{e\in E}\xi_{e}\left(\mathbf{P}_{\boldsymbol{\Gamma}}^{T}\right)_{e}\frac{\partial\mathcal{V}_{e}^{int}(\boldsymbol{\Gamma}_{e}(\mathbf{q}))}{\partial\boldsymbol{\Gamma}_{e}(\mathbf{q})}=\mathbf{P}_{\boldsymbol{\Gamma}}^{T}\mathbf{f}^{ext}_c\,.\label{eqn:QMHR}
\end{gather}

We now show that the hyper-reduced model \eqref{eqn:QMHR}, obtained
using EECSW, is also structure-preserving. Analogous to the previous discussion, the Hamiltonian associated to the hyper-reduced system becomes 
\begin{align*}
\tilde{\mathcal{E}}_{r}(\mathbf{q},\dot{\mathbf{q}}) & =\mathcal{\tilde{T}}_{r}(\dot{\mathbf{q}},\mathbf{q}) + \tilde{\mathcal{V}_{r}}^{int}(\mathbf{q})+\mathcal{V}^{ext}(\boldsymbol{\Gamma}(\mathbf{q}))\\
\implies\dot{\tilde{\mathcal{E}}}_{r}(\mathbf{q},\dot{\mathbf{q}},\ddot{\mathbf{q}}) & =\dot{\mathcal{\tilde{T}}}_{r}(\dot{\mathbf{q}},\mathbf{q})+\dot{\tilde{\mathcal{V}}}_{r}^{int}(\mathbf{q})+\dot{\mathcal{V}}^{ext}(\boldsymbol{\Gamma}(\mathbf{q}))\\
& =\dot{\mathbf{q}}^{T}\left(\sum_{e\in E} \xi_e \left(\mathbf{P}_{\boldsymbol{\Gamma}}^{T}\right)_{e} \mathbf{M}_e \ddot{\boldsymbol{\Gamma}}_e (\mathbf{q})+\sum_{e\in E}\xi_{e}\left(\mathbf{P}_{\boldsymbol{\Gamma}}^{T}\right)_{e}\frac{\partial\mathcal{V}_{e}^{int}(\boldsymbol{\Gamma}_{e}(\mathbf{q}))}{\partial\boldsymbol{\Gamma}_{e}(\mathbf{q})}-\mathbf{P}_{\boldsymbol{\Gamma}}^{T}\mathbf{f}^{ext}_c\right)\\
& =0\qquad(\text{Using (\ref{eqn:QMHR})}).
\end{align*}
This shows that hyper-reduced model \eqref{eqn:QMHR} also conserves energy and preserves the Lagrangian structure associated to the corresponding Hamilton's
principle. The numerical stability
of the hyper-reduced model is a direct consequence of this structure-preserving
property as noted in \cite{ECSW2}, as well.

\subsubsection{Dissipative, non-conservative systems}
To include dissipation effects which, typically, generate forces opposite
to the generalized velocity vector $\dot{\mathbf{u}}$ in the system,
a dissipation functional $\mathcal{D}$ can be defined, such that
the corresponding dissipative forces are given by 
\[
\mathbf{f}^{dis}(\dot{\mathbf{u}})=-\frac{\partial\mathcal{D}(\dot{\mathbf{u}})}{\partial\dot{\mathbf{u}}}.
\]
Assuming that the functional $ \mathcal{D} $ has the same degree of homogeneity, $ d $, in the generalized velocities ($ \dot{\mathbf{u}} $) for all elements in the mesh, we have 
\begin{equation}
\label{eq:disfun}
\dot{\mathbf{u}}^T \frac{\partial\mathcal{\mathcal{D}}(\dot{\mathbf{u}})}{\partial\dot{\mathbf{u}}} = \sum_{e=1}^{n_e} \dot{\mathbf{u}}^T_e \frac{\partial\mathcal{D}_e(\dot{\mathbf{u}}_e)}{\partial\dot{\mathbf{u}}_e} = d\sum_{e=1}^{n_e} \mathcal{D}_e(\dot{\mathbf{u}}_e),
\end{equation}
where $ \mathcal{D}_e $ is the usual notation for the contribution of element $ e $ towards the total dissipation $ \mathcal{D} $, which would be dependent on the nodal velocities $ \dot{\mathbf{u}}_e $ only. For the special case of linear viscous damping, i.e., $ d = 2 $, we have 
\[
\mathcal{\mathcal{D}}(\dot{\mathbf{u}})=\frac{1}{2}\dot{\mathbf{u}}^{T}\mathbf{C}\dot{\mathbf{u}}.
\]
Other values of $ d $ lead to different types of dissipation models, e.g., $ d = 1 $ leads to dry Coloumb friction and $ d=3 $ leads to a viscous drag in fluids. 

The Hamilton's principle in this non-conservative setting leads to
\begin{equation}
\label{eq:disHam}
\dot{\mathcal{T}}+\dot{\mathcal{V}}=\left(-\frac{\partial\mathcal{\mathcal{D}}(\dot{\mathbf{u}})}{\partial\dot{\mathbf{u}}}+\mathbf{f}^{ext}_{nc}\right)\cdot\dot{\mathbf{u}}=  \dot{\mathbf{u}}^T\mathbf{f}^{ext}_{nc} - d\sum_{e=1}^{n_e}  \mathcal{D}_e(\dot{\mathbf{u}}_e),
\end{equation}
where $ \mathbf{f}^{ext}_{nc} $  is a non-conservative, time-dependent external force. The corresponding Lagrange equations of motion can then be expressed, equivalently,
as 
\begin{equation}
\label{eq:ncHFM}
\mathbf{M}\ddot{\mathbf{u}}+\frac{\partial\mathcal{\mathcal{D}}(\dot{\mathbf{u}})}{\partial\dot{\mathbf{u}}}+\frac{\partial\mathcal{V}^{int}(\mathbf{u})}{\partial\mathbf{u}}=\mathbf{f}^{ext}_c+\mathbf{f}^{ext}_{nc}.
\end{equation}

Now, the nonlinear ROM associated to \eqref{eq:ncHFM} is given by
\begin{equation}
\label{eq:NLMdisred}
\mathbf{P}_{\boldsymbol{\Gamma}}^{T}\left[\mathbf{M}\ddot{\boldsymbol{\Gamma}}(\mathbf{q})+ \frac{\partial\mathcal{D}(\dot{\boldsymbol{\Gamma}}(\mathbf{q}))}{\partial\dot{\boldsymbol{\Gamma}}(\mathbf{q})} + \frac{\partial\mathcal{V}^{int}(\boldsymbol{\Gamma}(\mathbf{q}))}{\partial\boldsymbol{\Gamma}(\mathbf{q})} \right]=\mathbf{P}_{\boldsymbol{\Gamma}}^{T}[\mathbf{f}^{ext}_c + \mathbf{f}^{ext}_{nc}].
\end{equation}
The approximation of the additional nonlinear term in \eqref{eq:NLMdisred} using EECSW is given by \begin{equation}
\mathbf{P}_{\boldsymbol{\Gamma}}^{T}\frac{\partial\mathcal{D}(\dot{\boldsymbol{\Gamma}}(\mathbf{q}))}{\partial\dot{\boldsymbol{\Gamma}}(\mathbf{q})} = 
\sum_{e=1}^{n_e} \left(\mathbf{P}_{\boldsymbol{\Gamma}}^{T}\right)_e \frac{\partial\mathcal{D}_e\left(\dot{\boldsymbol{\Gamma}}_e(\mathbf{q})\right)}{\partial\dot{\boldsymbol{\Gamma}}_e(\mathbf{q})}\approx \frac{\partial\tilde{\mathcal{D}}_r(\dot{\mathbf{q}},\mathbf{q})}{\partial\mathbf{q}} =
\sum_{e\in E} \left(\mathbf{P}_{\boldsymbol{\Gamma}}^{T}\right)_e \frac{\partial\mathcal{D}_e\left(\dot{\boldsymbol{\Gamma}}_e(\mathbf{q})\right)}{\partial\dot{\boldsymbol{\Gamma}}_e(\mathbf{q})}\,,
\end{equation}
where
\begin{equation}
\label{eq:disfunHR}
\tilde{\mathcal{D}}_r(\dot{\mathbf{q}},\mathbf{q}) = \sum_{e\in E}\xi_e \mathcal{D}_e(\dot{\boldsymbol{\Gamma}}_e(\mathbf{q})),
\end{equation}
can be thought of as the dissipation associated to the hyper-reduced model, given in final form as
\begin{equation}
\label{eq:HROMdis}
\sum_{e\in E} \xi_e \left(\mathbf{P}_{\boldsymbol{\Gamma}}^{T}\right)_{e} \left[ \mathbf{M}_e \ddot{\boldsymbol{\Gamma}}_e (\mathbf{q}) + \frac{\partial\mathcal{D}_e\left(\dot{\boldsymbol{\Gamma}}_e(\mathbf{q})\right)}{\partial\dot{\boldsymbol{\Gamma}}_e(\mathbf{q})} + \frac{\partial\mathcal{V}_{e}^{int}(\boldsymbol{\Gamma}_{e}(\mathbf{q}))}{\partial\boldsymbol{\Gamma}_{e}(\mathbf{q})} \right] =\mathbf{P}_{\boldsymbol{\Gamma}}^{T}[\mathbf{f}^{ext}_c + \mathbf{f}^{ext}_{nc}].
\end{equation}

Following the same procedure as shown above for conservative systems, it is easy to see that the Lagrangian structure is preserved using EECSW for non-conservative systems such as \eqref{eq:HROMdis}, as well. Indeed, the rate of change of the total energy $ \tilde{\mathcal{E}}_r $ associated to the hyper-reduced model \eqref{eq:HROMdis} is given by 
\begin{align*}
\dot{\tilde{\mathcal{E}}}_{r}(\mathbf{q},\dot{\mathbf{q}},\ddot{\mathbf{q}}) & =\dot{\mathcal{\tilde{T}}}_{r}(\dot{\mathbf{q}},\mathbf{q})+\dot{\tilde{\mathcal{V}}}_{r}^{int}(\mathbf{q})+\dot{\mathcal{V}}^{ext}(\boldsymbol{\Gamma}(\mathbf{q})) - \frac{d}{dt}\left(\boldsymbol{\Gamma}(\mathbf{q}) \cdot \mathbf{f}^{ext}_{nc}\right)\\
& = \dot{\mathbf{q}}^{T}\left(\sum_{e\in E} \xi_e \left(\mathbf{P}_{\boldsymbol{\Gamma}}^{T}\right)_{e} \left[\mathbf{M}_e \ddot{\boldsymbol{\Gamma}}_e (\mathbf{q})+\frac{\partial\mathcal{V}_{e}^{int}(\boldsymbol{\Gamma}_{e}(\mathbf{q}))}{\partial\boldsymbol{\Gamma}_{e}(\mathbf{q})}\right]-\mathbf{P}_{\boldsymbol{\Gamma}}^{T}[\mathbf{f}^{ext}_c+\mathbf{f}^{ext}_{nc}]\right) - \boldsymbol{\Gamma}(\mathbf{q})\cdot\dot{\mathbf{f}}^{ext}_{nc} \\
& = -\sum_{e\in E} \xi_e \dot{\mathbf{q}}^{T}\left(\mathbf{P}_{\boldsymbol{\Gamma}}^{T}\right)_{e} \frac{\partial\mathcal{D}_e\left(\dot{\boldsymbol{\Gamma}}_e(\mathbf{q})\right)}{\partial\dot{\boldsymbol{\Gamma}}_e(\mathbf{q})} - \boldsymbol{\Gamma}(\mathbf{q})\cdot\dot{\mathbf{f}}^{ext}_{nc} \qquad(\text{Using (\ref{eq:HROMdis})}) \\
& = -\sum_{e\in E} \xi_e \left(\dot{\boldsymbol{\Gamma}}_e(\mathbf{q})\right)^T \frac{\partial\mathcal{D}_e\left(\dot{\boldsymbol{\Gamma}}_e(\mathbf{q})\right)}{\partial\dot{\boldsymbol{\Gamma}}_e(\mathbf{q})} - \boldsymbol{\Gamma}(\mathbf{q})\cdot\dot{\mathbf{f}}^{ext}_{nc}  \\
& = -d \tilde{\mathcal{D}}_r(\dot{\mathbf{q}},\mathbf{q}) - \boldsymbol{\Gamma}(\mathbf{q})\cdot\dot{\mathbf{f}}^{ext}_{nc} \qquad(\text{Using \eqref{eq:disfun}, \eqref{eq:disfunHR}})
\end{align*}
\begin{equation}
\label{eq:dis_proof}
\implies \dot{\mathcal{\tilde{T}}}_{r}(\dot{\mathbf{q}},\mathbf{q})+\dot{\tilde{\mathcal{V}}}_{r}^{int}(\mathbf{q})+\dot{\mathcal{V}}^{ext}(\boldsymbol{\Gamma}(\mathbf{q})) = -d \tilde{\mathcal{D}}_r(\dot{\mathbf{q}},\mathbf{q}) + \left(\dot{\boldsymbol{\Gamma}}(\mathbf{q})\right)^T \mathbf{f}^{ext}_{nc}.
\end{equation}
When compared with \eqref{eq:disHam} and \eqref{eq:disfunHR}, Equation \eqref{eq:dis_proof} proves that EECSW preserves the Lagrangian structure associated to the Hamiltonian principle for a non-conservative system, as well. 

The conditional/unconditional numerical stability properties of any chosen time integration scheme are carried over to the hyper-reduced model as a direct consequence of the energy-conserving/ structure-preserving properties of EECSW. We refer the reader to Section 4.3 in \cite{ECSW2} for the discussion on how these properties lead to numerical stability.

\section{Numerical results}

\subsection{Setup}

\label{subsec:resultsSetup}

To test and illustrate the performance of the EECSW method, proposed
here, we require a high-dimensional system which can be reduced using
a suitable nonlinear mapping. To this end, we use the quadratic manifold,
introduced in \cite{qm}, for model reduction. In the quadratic manifold
approach, the full set of unknowns $\mathbf{u}$ are mapped to a lower
dimensional set of variables $\mathbf{q}$ using a quadratic mapping
as

\begin{equation}
\mathbf{u}\approx\boldsymbol{\Gamma}(\mathbf{q}):=\boldsymbol{\Phi}\mathbf{q}+\frac{1}{2}\boldsymbol{\Omega}:\left(\mathbf{q}\otimes\mathbf{q}\right),\label{eq:QM}
\end{equation}
where the columns of $\boldsymbol{\Phi}\in\mathbb{R}^{n\times M}$
contains $M$ significant vibration modes (VMs) extracted at equilibrium
$\mathbf{u}=\mathbf{0}$, and $\boldsymbol{\Omega}\in\mathbb{R}^{n\times M\times M}$
is a third-order tensor containing the corresponding static modal
derivatives (SMDs). The symbol $\otimes$ denotes the dyatic product,
while $:$ indicates contraction over the last two indices \footnote{The mapping (\ref{eq:QM}) can be written using the Einstein summation
	convention in the indicial notation as: $\Gamma_{I}=\Phi_{Ii}q_{i}+\frac{1}{2}\Omega_{Iij}q_{i}q_{j}\quad I\in\{1,\dots,n\},\quad i,j\in\{1,\dots,M\}$}.

For consistency, we use the same examples used in \cite{qm}: Model-I,
a flat plate, simply-supported on two opposite sides, and Model-II,
a NACA-airfoil-based wing model. The details of both models are shown
in Figures \ref{fig:Model1} and \ref{fig:Model2}. Time integration is performed using the implicit Newmark scheme with a time step constant across all techniques for a fair comparison. Both structures
are discretized using flat, triangular-shell elements with 6 DOFs
per node, i.e., 18 DOFs per element. For both models, the accuracy
of the results is compared to the corresponding full nonlinear solutions
using a mass-normalized global relative error (GRE) measure, defined
as 
\[
GRE_{M}=\dfrac{\sqrt{\sum\limits _{t\in S}(\mathbf{u}(t)-\tilde{\mathbf{u}}(t))^{T}\mathbf{M}(\mathbf{u}(t)-\tilde{\mathbf{u}}(t))}}{\sqrt{\sum\limits _{t\in S}\mathbf{u}(t)^{T}\mathbf{M}\mathbf{u}(t)}}\times100,
\]
where $\mathbf{u}(t)\in\mathbb{R}^{n}$ is the vector of generalized
displacements at the time $t$, obtained from the HFM solution, $\tilde{\mathbf{u}}(t)\in\mathbb{R}^{n}$
is the solution based on the (hyper)reduced model, and $S$ is the
set of time instants at which the error is recorded. The mass matrix
$\mathbf{M}$ provides a relevant normalization for the generalized
displacements, which could be a combination of physical displacements
and rotations, as is the case in the shell models shown here.

The computations were performed on the Euler cluster of ETH Zürich.
Since the success of reduction techniques is often reported in terms
of savings in simulation time, we define an online speedup $S^{\star}$
computed according to the following simple formula: 
\[
S^{\star}=\frac{T_{full}}{T_{sim}^{\star}},
\]
where $T_{full}$ and $T_{sim}^{\star}$ represent the CPU time taken
during the time integration of $\textit{full}$ and (hyper)reduced
solution respectively. The superscript $\star$ denotes the reduction
technique being used.

We compare the hyper-reduction using ECSW (where the ROMs are obtained
from linear mappings), with the EECSW method. For both ECSW and EECSW,
the convergence tolerance $\tau$ for finding a sparse solution to
the NNLS problem (\ref{eq:NNLS}), as required in Algorithm \ref{alg:sNNLS},
is chosen to be 0.01. This is within the practically recommended range
as reported in \cite{ECSW1}. A total of $n_{t}=200$ training vectors,
chosen uniformly from the solution time-span, are used in sparse NNLS
routine. The following ROMs are considered in our comparison: 
\begin{itemize}
	\item \emph{Linear-mapping}-\emph{based (Hyper-)Reduction}: 
	\begin{description}
		\item [{POD:}] Here, the first most significant $m$ Proper Orthogonal
		Decomposition (POD) modes obtained from the HFM run are used to construct
		$\mathbf{V}$. 
		\item [{ECSW-POD:}] Hyper-reduction using ECSW is performed, for the ROM
		obtained in the POD approach. This is the original ECSW implementation,
		as proposed in \cite{ECSW1}. 
	\end{description}
	\item \emph{Quadratic-manifold-based (Hyper-)Reduction}: 
	\begin{description}
		\item [{QM:}] Here, the quadratic mapping (\ref{eq:QM}) is used to construct
		the ROM as given by (\ref{eq:NLROM}). 
		\item [{EECSW-QM:}] The ROM obtained in QM approach using the quadratic
		manifold is hyper-reduced using the EECSW. 
	\end{description}
\end{itemize}

\subsection{Flat Structure}
\begin{figure}[H]
	\subfloat[]{\centering{}\includegraphics[width=0.48\linewidth]{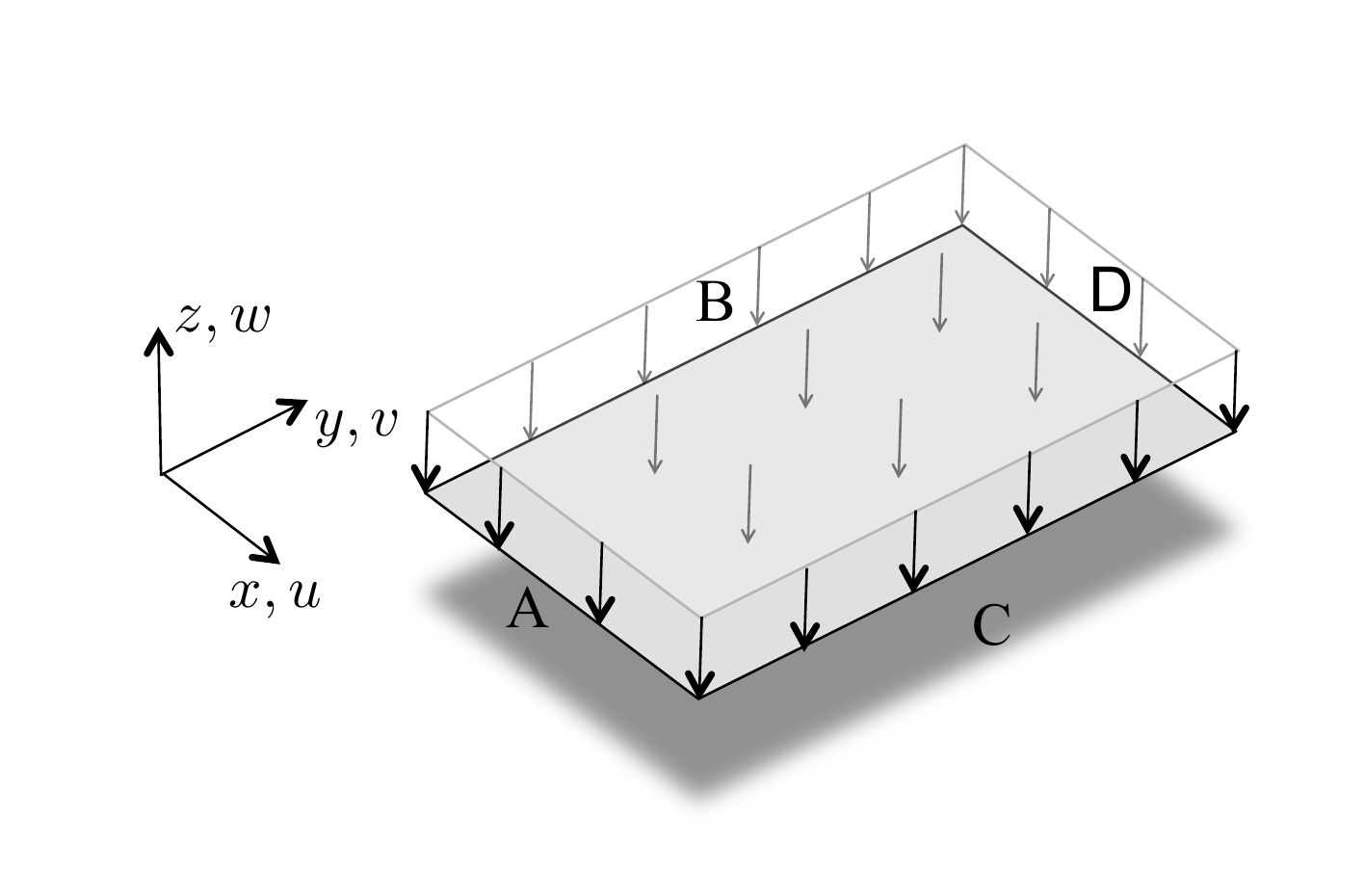}}\hfill{}\subfloat[]{\centering{}\includegraphics[width=0.48\linewidth]{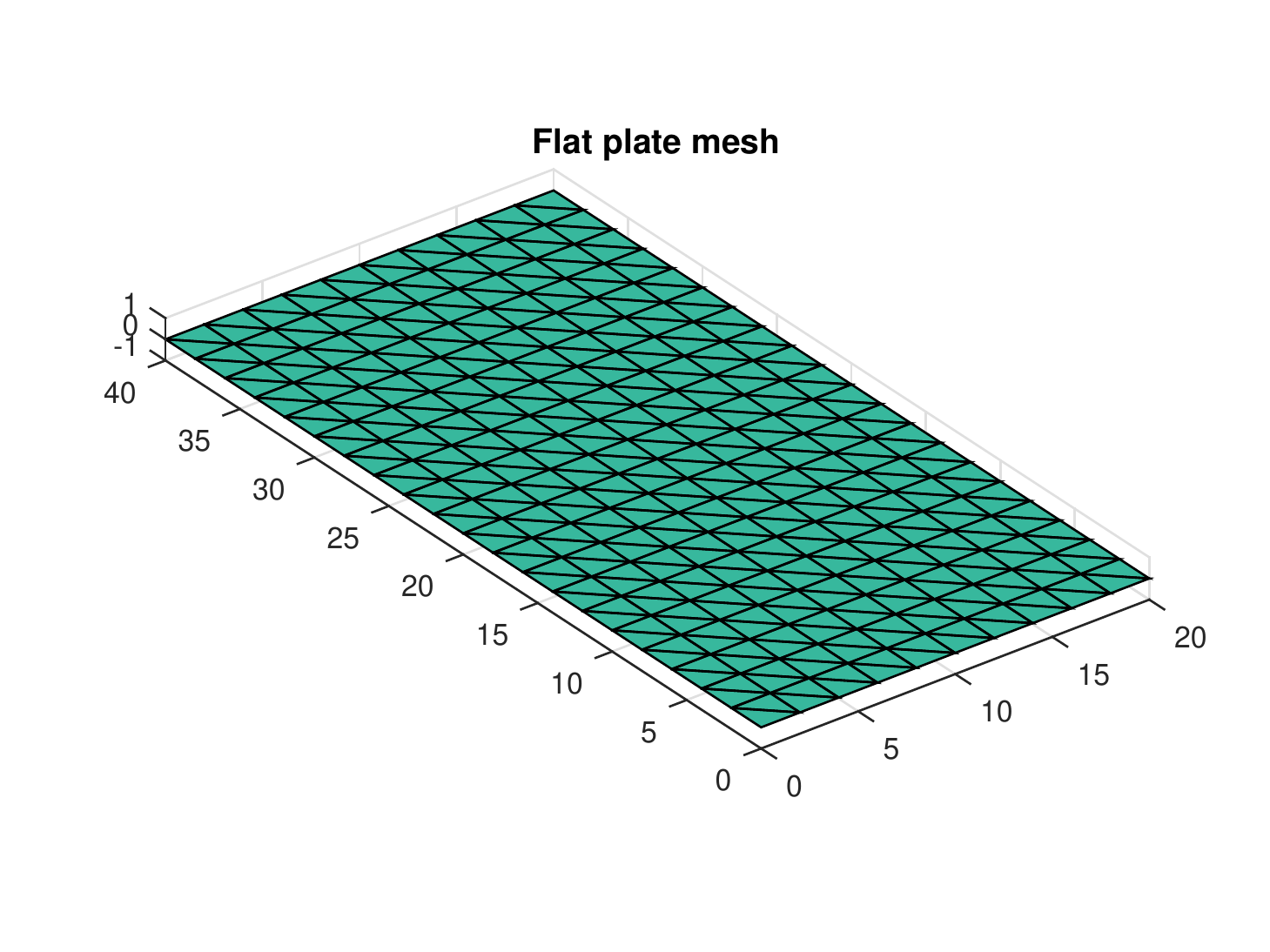}}
	
	\caption{\label{fig:Model1} \textbf{Model - I:} (a) Flat plate example - simply
		supported on sides $A$ and $D$. The plate is $L=40$ mm long, $H=20$
		mm wide, $t=0.8$ mm thick. The Young Modulus is $E=70$ GPa, the
		Poisson's ratio is $\nu=0.33$, and the density is $\rho=2700$ Kg/m$^{3}$.
		A uniform pressure is applied on the plate, according to the time
		history $p(t)=P\sin(\omega t)$, where $P=1$ N/mm$^{2}$ and $\omega=2.097\times10^{4}$
		rad/s. (b) The curved plate is discretized using flat, triangular
		shell elements. The FE mesh contains 400 elements and 1386 DOFs. The
		two opposite sides parallel to the $y$-axis are simply supported,
		a uniform pressure is applied on the curved surface, according to
		the time history. }
\end{figure}
\begin{figure}[H]
	\centering{}\subfloat[]{\includegraphics[width=0.49\linewidth]{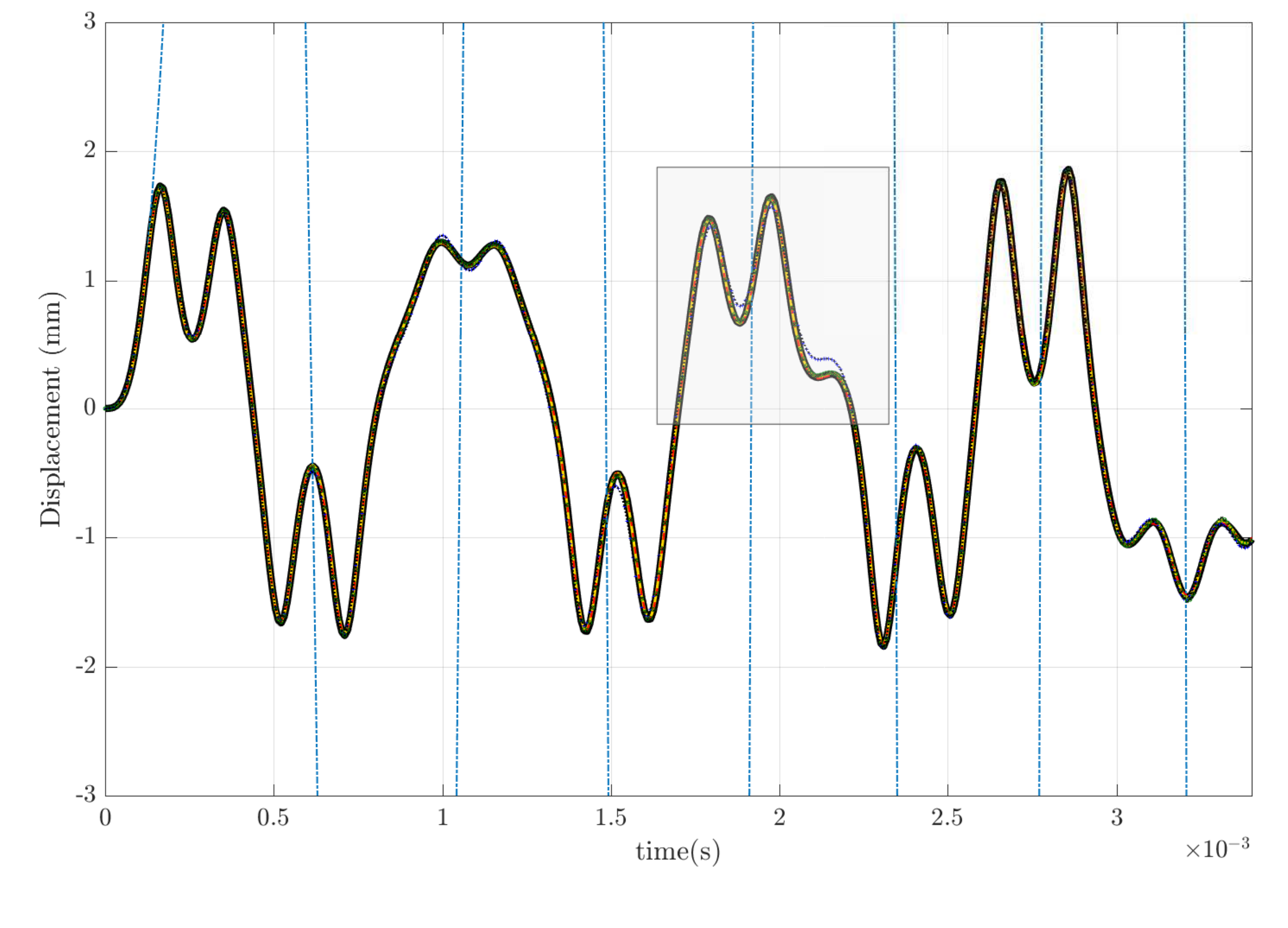}}\hfill{}\subfloat[]{\includegraphics[width=0.49\linewidth]{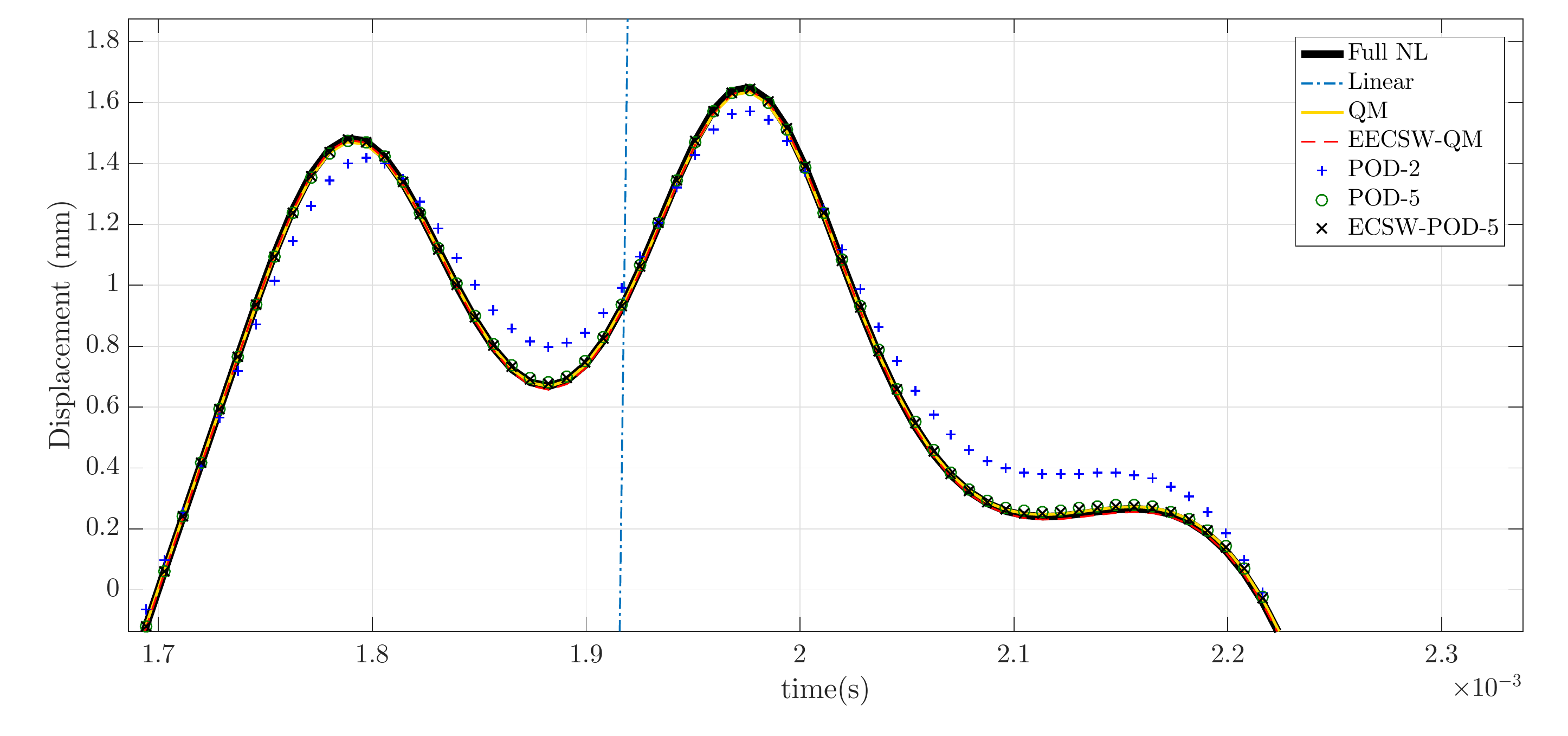}}
	
	\caption{\label{fig:Model1-resp} The linear solution (dashed-blue) grows quickly
		to high amplitudes for the choice of resonant loading; the full solution
		(thick black) shows much lower amplitudes due to the geometrical nonlinearities;
		the ROM obtained using quadratic manifold (yellow) and hyper-reduced
		solution for this ROM using EECSW (dashed-red) capture the full solution
		accurately. Subfigure (b) highlights the inaccuracy of the POD ROM with $m=2$.}
\end{figure}

We consider the rectangular flat plate structure shown in Figure \ref{fig:Model1}.
A uniform pressure distribution is chosen to act normal to the plate
surface as the external load. A time varying amplitude (load function)
is used given by

\begin{gather}
\mathbf{g}(t)=p(t)\mathbf{l},\nonumber \\
p(t)=\sin(\omega t),\label{eqn:loadfactor}
\end{gather}
where $\mathbf{l}$ is a constant load vector corresponding to a uniform
pressure distribution of 1 N/mm$^{2}$. Here, $p(t)$, termed as the
dynamic load function, determines the time-dependency of the external
load. The results are shown for a periodic choice for $p(t)$ as given
by (\ref{eqn:loadfactor}), where $\omega$ is chosen to be the first
natural frequency of the linearized system. Due to the resonant nature
of the dynamics, it is easy to excite geometrically nonlinear behavior
in the system, as shown by Figure \ref{fig:Model1-resp}. Furthermore,
Figure \ref{fig:Model1-resp} shows the that the full nonlinear solution
is reduced using the QM technique and, subsequently, hyper-reduced
using EECSW with good accuracy.

It can also be confirmed from the global relative error $GRE_{M}$
and speed-up $S^{\star}$, shown in Table \ref{tab:Model-I-Results},
that the QM based nonlinear ROM reproduces the results of Model-I
with good accuracy but, as expected, does not deliver a good speed-up.
The EECSW-based hyper-reduction, on the other hand, gives results
with a $GRE_{M}$ of 1.37\% but delivers a much higher speed-up of about 100. Note also that a linear-mapping-based ROM with POD of the same size ($m=2$) as the adopted nonlinear mapping does not deliver a good accuracy.

\begin{table}[H]
	\begin{centering}
		\begin{tabular}{|>{\raggedright}m{2.7cm}|>{\centering}p{1.25cm}|>{\centering}p{1.6cm}|c|c|}
			\hline 
			Reduction Technique  & Basis size ($m$)  & \# elements  & $GRE_{M}$(\%)  & $S^{\star}$\tabularnewline
			\hline 
			\hline 
			POD & 2 & 400 & 5.62 & 2.96\tabularnewline
			\hline 
			POD  & 5  & 400  & 1.81  & 2.84\tabularnewline
			\hline 
			ECSW-POD  & 5  & 12  & 2.52  & 77.7\tabularnewline
			\hline 
			\hline 
			QM  & 2  & 400  & 1.32  & 3.03\tabularnewline
			\hline 
			\textbf{EECSW-QM}  & 2  & 6  & 1.37  & 99.93\tabularnewline
			\hline 
		\end{tabular}
		\par\end{centering}
	\caption{\label{tab:Model-I-Results}Comparison of global relative error $GRE_{M}$
		and speed-ups in Model-I for linear-mapping-based ROMs and their hyper-reduced
		counterparts on one hand, and for quadratic-manifold-based ROMs and
		their hyper-reduced counterparts using EECSW on the other hand. Time
		for a full solution run $T_{full}=74.4$ s.}
\end{table}

\subsection{Wing Structure}
\begin{figure}[H]
	\subfloat[]{\centering{}\includegraphics[width=0.48\linewidth]{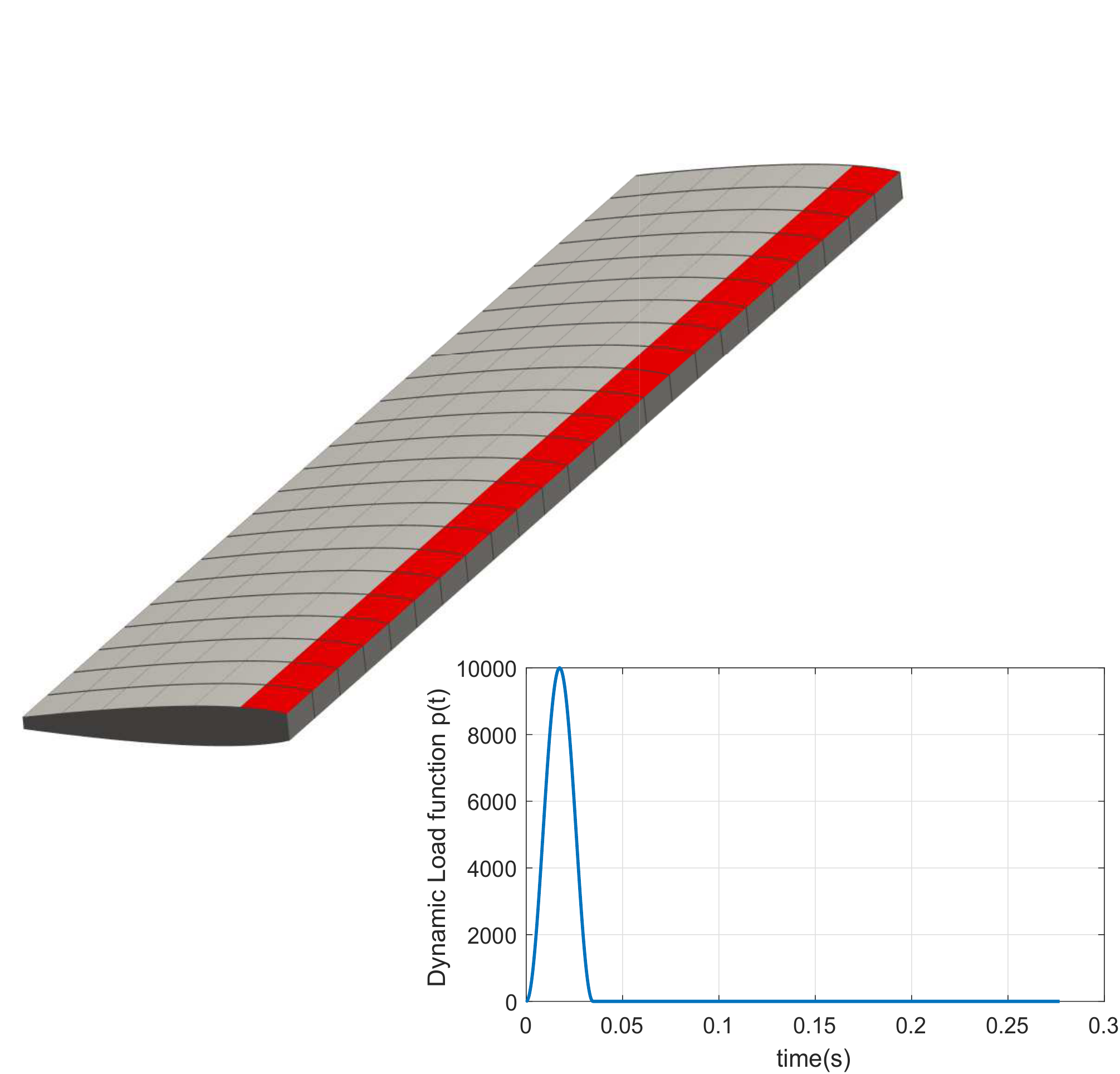}}\hfill{}\subfloat[]{\centering{}\includegraphics[width=0.48\linewidth]{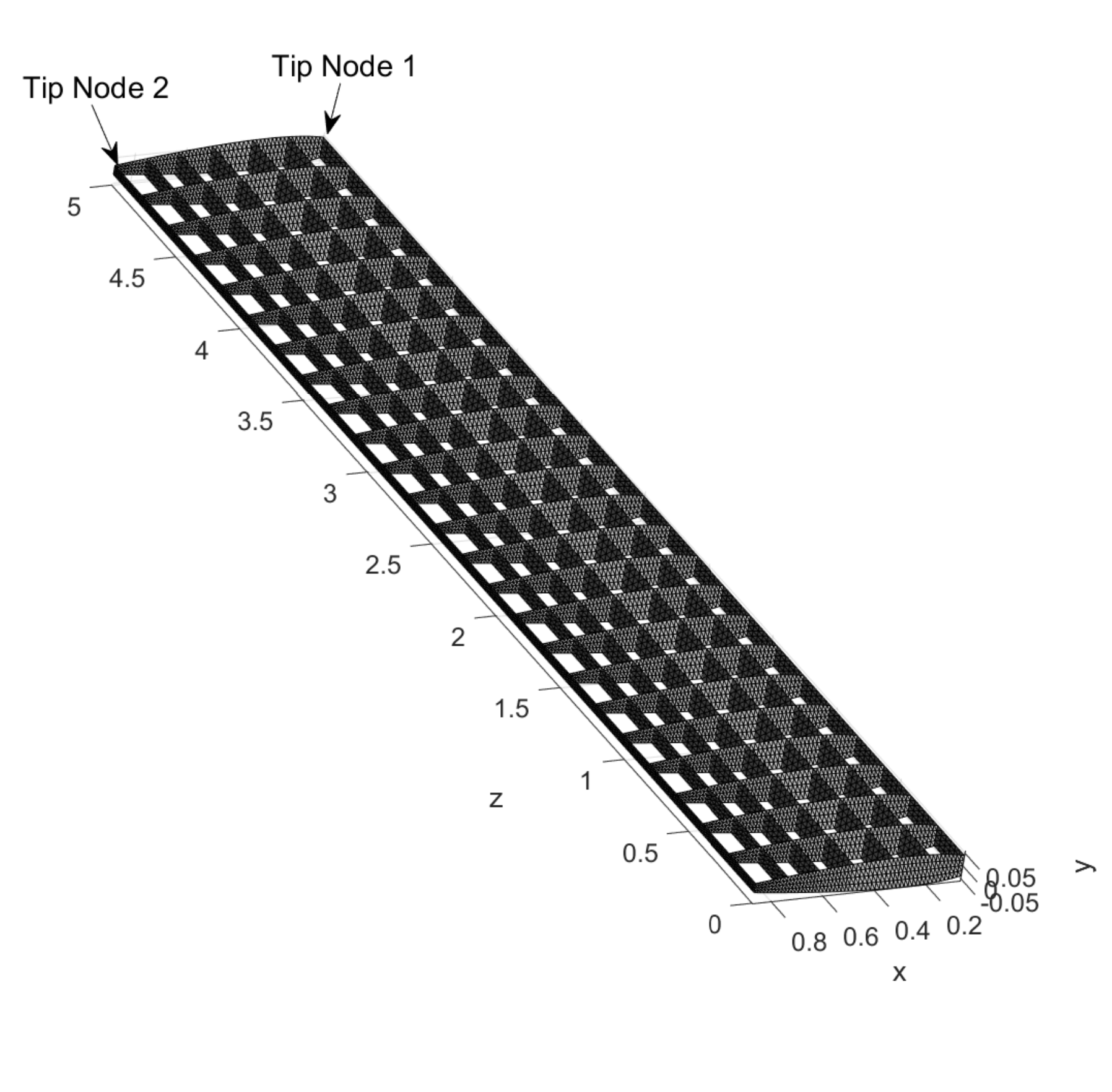}}
	
	\caption{\label{fig:Model2} \textbf{Model - II}: (a) A wing structure with
		NACA 0012 airfoil (length($L$) = 5 m, Width($W$) $\approx$ 0.9
		m, Height($H$) = 0.1 m ) stiffened with ribs along the longitudinal
		and lateral direction. The Young Modulus is $E=70$ GPa, the Poisson's
		ratio is $\nu=0.33$, and the density is $\rho=2700$ Kg/m$^{3}$.
		The wing is cantilevered at one end. Uniform pressure is applied on
		the highlighted area, with a pulse load as given by (\ref{eq:Model2_load});
		(b) The structure is meshed with triangular flat shell elements with
		6 DOFs per node and each with a thickness of 1.5 mm. The mesh contains
		$n=135770$ DOFs, $n_{e}=49968$ elements. For illustration purposes,
		the skin panels are removed, and mesh is shown.}
\end{figure}
\begin{figure}[H]
	\centering{}
	\subfloat[]{\includegraphics[width=0.52\linewidth]{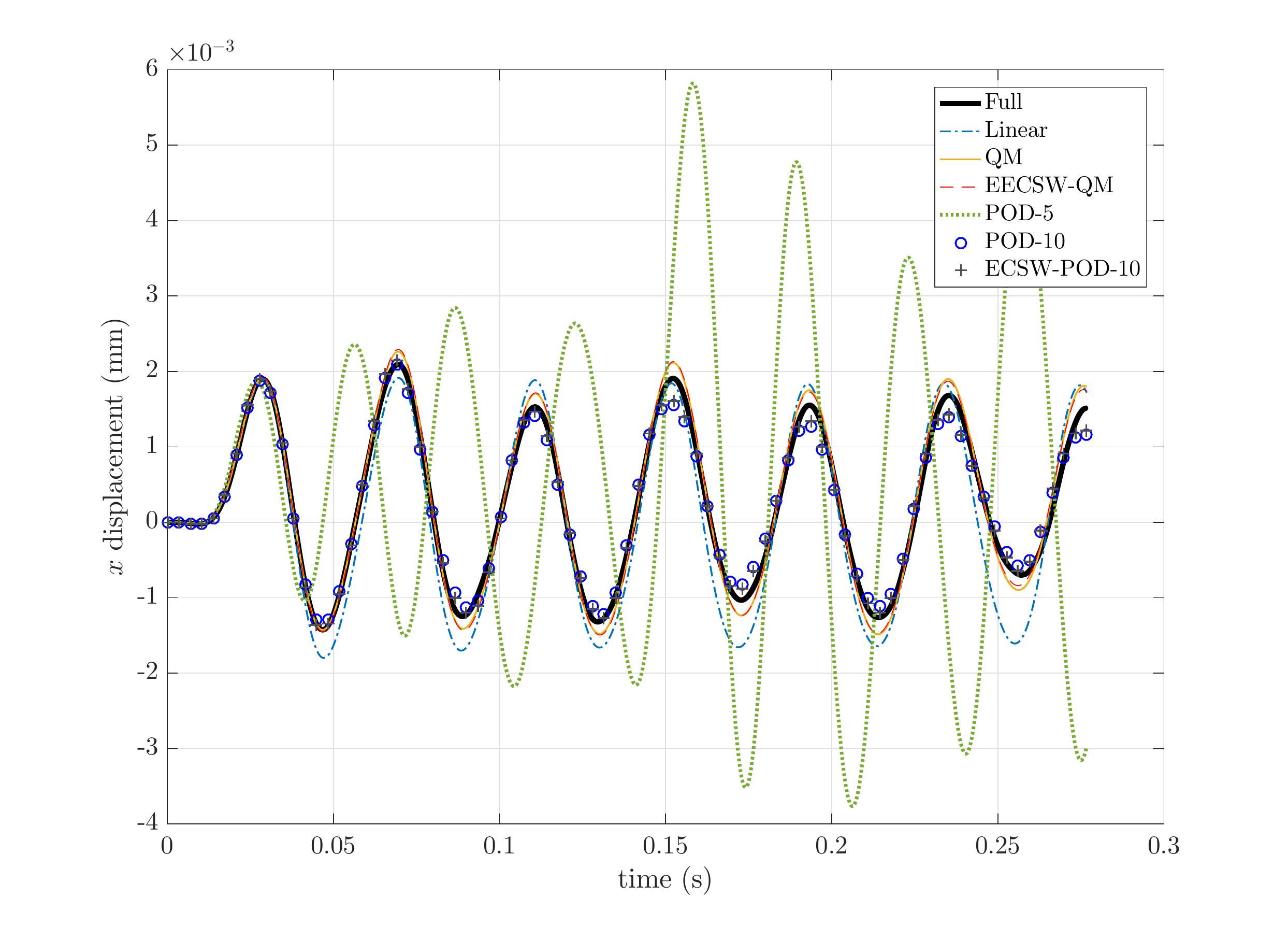}}
	\subfloat[]{\centering{}\includegraphics[width=0.52\linewidth]{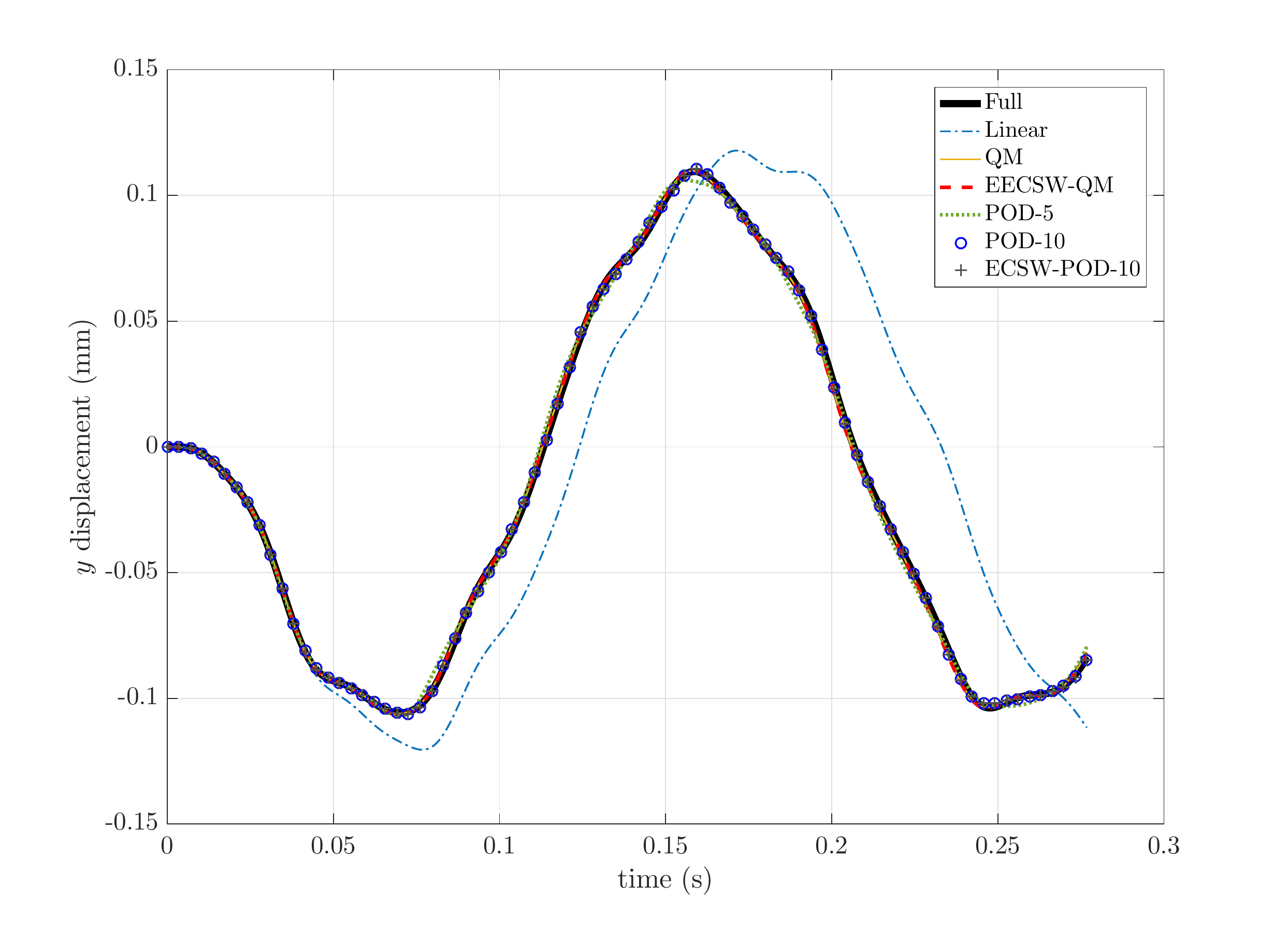}}
	
	\subfloat[]{\centering{}\includegraphics[width=0.52\linewidth]{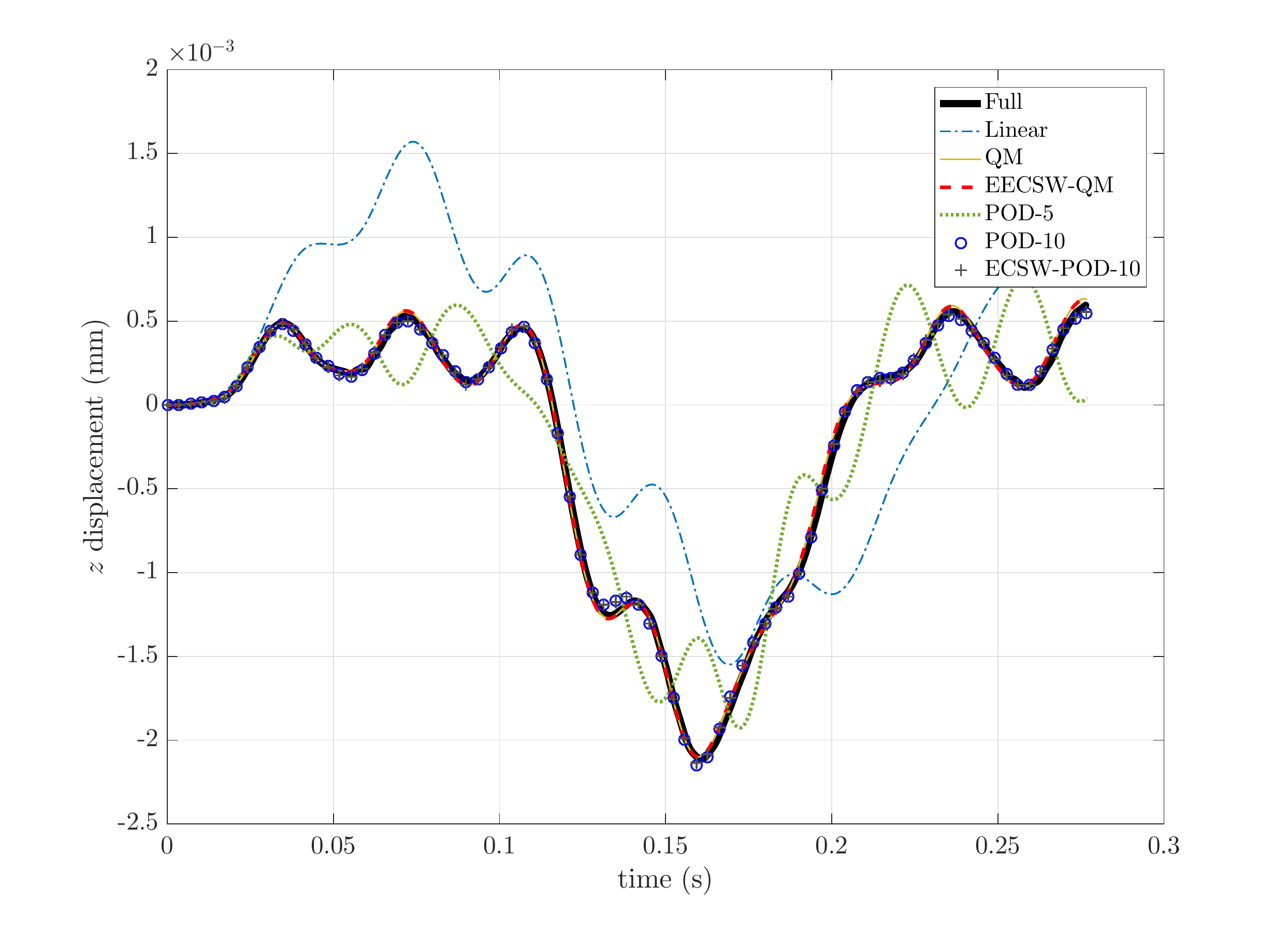}}
	
	\caption{\label{fig:Model2-resp} (a), (b), (c): Comparison of full, linear and various reduced responses at tip node 2 in $x,y$ and $z$ directions. Note that the QM-based
		reduced solution (with 5 reduced unknowns) accurately captures the
		full system response, where as the POD-based ROM with 5 modes fails
		to reproduce the nonlinear system response (POD-5). }
\end{figure}

For the application of the proposed extension of EECSW to more realistic models, we consider the model of a NACA-airfoil-based
wing structure, introduced in \cite{qm}. This model, referred to
as Model-II, contains truly high number of DOFs ($n=135770$), thereby,
allowing for the appreciation of obtained accuracy and computational
speed-ups. We simulate the response of the structure to a low frequency
pulse load, applied as a spatially uniform pressure load on the highlighted
area on the structure skin (cf. Figure \ref{fig:Model2}). The dynamic
load function, shown in Figure \ref{fig:Model2-resp}a, is given as
\begin{equation}
p(t)=A\sin^{2}(\omega t)\left[H(t)-H\left(\frac{\pi}{\omega}-t\right)\right],\label{eq:Model2_load}
\end{equation}
where $H(t)$ is the Heaviside function and $\omega$ chosen as the
mean of the first and second natural frequency of vibration. We refer
the reader to \cite{qm} for the detailed linearized, full nonlinear
and reduced nonlinear response plots of this model.

Figure \ref{fig:Model2-resp}b and Table \ref{tab:Model-II-Results}
show that the hyper-reduction using EECSW for the quadratic-manifold
based ROM reproduce the full system response with a very good accuracy
($GRE_{M}\approx1.71\%$) and is more than a thousand times faster
than the full response computation. As expected, we obtain a much
higher speed-up during hyper-reduction in comparison with Model-I.
The speed-up without hyper-reduction, however, are fairly similar
between the two models, despite the striking difference in the number
of DOFs. Thus, the numerical results confirm the need and effectiveness
of EECSW for hyper-reduction of large nonlinear systems over nonlinear
manifolds.

Figure \ref{fig:Model2-resp} and Table \ref{tab:Model-II-Results}
compare the response of Model-II to the applied load. Figure \ref{fig:Model2-resp}
further shows that the response of the full system is captured more
effectively using the quadratic manifold as compared to the POD with
the same number of unknowns (POD-5). This underscores the importance
of nonlinear manifolds in model reduction of nonlinear systems. In
terms of speed as well, we see in Table \ref{tab:Model-II-Results}
that the EECSW method performs better than its linear counterpart.

\begin{table}[H]
	\begin{centering}
		\begin{tabular}{|>{\raggedright}m{2.7cm}|>{\centering}p{1.25cm}|>{\centering}p{1.6cm}|c|>{\centering}p{1cm}|}
			\hline 
			Reduction Technique  & Basis size ($m$)  & \# elements  & $GRE_{M}$(\%)  & $S^{\star}$\tabularnewline
			\hline 
			\hline 
			POD  & 5  & 49968  & 5.01  & 3.07\tabularnewline
			\hline 
			POD  & 10  & 49968  & 0.98 & 2.77\tabularnewline
			\hline 
			ECSW-POD  & 10  & 152  & 1.05  & 794.6\tabularnewline
			\hline 
		\end{tabular}
		\par\end{centering}
	\centering{}\vspace{0.5mm}
	\begin{tabular}{|>{\raggedright}m{2.7cm}|>{\centering}p{1.25cm}|>{\centering}p{1.6cm}|>{\centering}p{1.68cm}|>{\centering}p{1cm}|}
		\hline 
		QM  & 5  & 49968  & 1.65  & 2.92\tabularnewline
		\hline 
		EECSW-QM  & 5  & 44  & 1.71  & 1098\tabularnewline
		\hline 
	\end{tabular}\caption{\label{tab:Model-II-Results}Starting with a linear modal superposition
	with $M=5$ modes (first 5 VMs), reduction techniques as described
	in Section \ref{subsec:resultsSetup} are formulated. Global Relative
	Error and speed-ups	in Model-II for these reduction techniques are tabulated. A total
	of $n_{t}=200$ training vectors, chosen uniformly from the solution
	time-span, are used to setup hyper-reduction in ECSW-POD and EECSW-QM. Time for
	a full solution run $T_{full}=3.808\times10^{4}$ s.}
\end{table}
It is interesting that in this case, EECSW-QM ends up sampling a much smaller set of elements as compared to its linear counterpart EECSW-POD, leading to a higher speed-up. This is not to say that nonlinear-manifold-based hyper-reduction would always outperform its linear mapping counterpart. In fact, for a given basis size ($ m $) and same number of elements in the sampled mesh, the EECSW method equipped with nonlinear mapping can be expected to be slower that ECSW-POD approach on two accounts:
\begin{enumerate}
	\item The additional convective terms in the nonlinear-mapping-based ROM \eqref{eq:NLROM2} adds to the computational burden in comparison with its linear counterpart \eqref{eq:GalerkinROM}
	\item The evaluation of the nonlinear mapping (quadratic in this case), is expected to be more expensive that a linear mapping, which boils down to a standard matrix-vector product. Though MATLAB is extremely efficient at performing such matrix-vector products, and hence, in the evaluation of linear mappings, the same cannot be said about our sub-optimal implementation for evaluation of the quadratic mapping at hand.	
\end{enumerate}
However, for large nonlinear systems, we expect nonlinear manifolds would require much lesser variables to capture the nonlinear behavior in comparison with linear mappings in order to achieve the desired accuracy (cf. discussion in Introduction). We have attempted to demonstrate this using a simple quadratic manifold on a large system, where only 5 unknowns were required to match the accuracy obtained using 10 POD modes (cf. Table \ref{tab:Model-II-Results} and Figure \ref{fig:Model2-resp}). In our experience, a smaller size reduced system usually leads to a smaller size of reduced mesh during training phase as well. This is intuitively clear since the sNNLS problem \ref{eq:NNLS} has fewer constraints to satisfy. Thus, despite the extra costs mentioned in the two points above, EECSW is capable of producing better computational speed-up when equipped with the proper nonlinear manifold (as also seen in Tables \ref{tab:Model-I-Results} and \ref{tab:Model-II-Results}). 



\section{Conclusion}
Nonlinear behavior of large structural systems can often be more effectively reduced using nonlinear manifolds in comparison with linear mappings. Though nonlinear manifolds help in reducing the number of unknowns in a large system, hyper-reduction is further needed to achieve the corresponding computational speed up using such ROMs. In this work, we extended the recently developed ECSW technique to allow for hyper-reduction using smooth nonlinear mappings into the reduced system of unknowns. In doing so, we also showed that the ROM obtained by projection onto the tangent space of the manifold, leads to the preservation of structure and stability properties of the
ECSW method. 

We demonstrated the EECSW method on two structural dynamics examples
by hyper-reduction of a ROM based on the quadratic manifold, whereby
better speed and accuracy were obtained using the EECSW method in
comparison with ECSW-POD based hyper-reduced models. Since these models
were based on von-Karman nonlinearities, the degree of nonlinearity
in the examples is expected to be much lower than the nominal thin-walled
structures featuring geometrically nonlinear behavior. A quadratic manifold was, therefore, effective in capturing the nonlinear behavior effectively for these structures. However, nonlinear manifolds -- possibly constructed from solution databases -- would play a crucial
role in the effective reduction of models with higher degree of nonlinearities (cf. \cite{Millan2013}). EECSW in such cases
is expected to play a vital role during the hyper-reduction stage
by providing the necessary computational speed-ups. 

Furthermore, it is easy to see that the proposed extension is capable of handling parameter-dependence in an analogous manner as the original ECSW. However, before applying hyper-reduction using EECSW, we require parameter-dependent nonlinear manifolds in construction of such ROMs. This is certainly not as straight forward as in the linear mapping context, where reduction bases are sometimes even {linearly} interpolated. Nonetheless, in data-based approaches, parameter dependence can be included into the manifold learning phase. In particular, the parameters can be explicitly modelled as the known features, apart from the reduced variables which would be identified in the manifold learning procedure. This would directly result in parameter-dependent nonlinear manifolds obtained from data. This underscores the need for manifold learning in model reduction with parameter dependence. Thus, using the EECSW on ROMs based on nonlinear manifolds constructed using manifold-learning techniques forms an essential part of our current and future efforts. 

\paragraph*{Acknowledgements}

The authors acknowledge the support of the Air Force Office of Scientific
Research, Air Force Material Command, USAF under Award No.FA9550-16-1-0096.
We also thank Ludovic Renson for pointing us to the MATLAB-based tensor
product toolbox TPROD.

\end{document}